\documentclass[12pt,preprint]{aastex}

\slugcomment{}

\shorttitle{Magnetic Fields in Giant Molecular Clouds}
\shortauthors{H. Li et al.}

\begin{document}


\title{Results of SPARO 2003: Mapping Magnetic Fields in Giant Molecular Clouds} 

\author{H. Li, G. S. Griffin, M. Krejny, and G. Novak }
\affil{Department of Physics and Astronomy, Northwestern University, Evanston, IL 60208}
\email{h-li5@northwestern.edu}

\author{R. F. Loewenstein, and M. G. Newcomb}
\affil{Yerkes Observatory, Williams Bay, WI 53191}

\author{P. G. Calisse }
\affil{School of Physics and Astronomy, Cardiff University, Wales, U.K.}
\and

\author{D. T. Chuss}
\affil{NASA - Goddard Space Flight Center, Greenbelt, MD 20771}




\begin{abstract}
We present results from the Austral Winter 2003 observing campaign of
\mbox{SPARO}, a 450 \micron\ polarimeter used with a two-meter telescope at South
Pole.  We mapped large-scale magnetic fields in four Giant Molecular
Clouds (GMCs) in the Galactic disk:  NGC 6334, the Carina Nebula,
G333.6-0.2 and G331.5-0.1.  We find a statistically significant
correlation of the inferred field directions with the orientation of the
Galactic plane.  Specifically, three of the four GMCs (NGC 6334 is the
exception) have mean field directions that are within 15\degr\ of the
plane.  The simplest interpretation is that the field direction tends to
be preserved during the process of GMC formation.  We have also carried
out an analysis of published optical polarimetry data.  For the closest of
the SPARO GMCs, NGC 6334, we can compare the field direction in the cloud
as measured by SPARO with the field direction in a larger region (several
hundred pc) surrounding the cloud, as determined from optical polarimetry.  
For purposes of comparison, we also use optical polarimetry to determine
field directions for 9-10 other regions of similar size.  We find that the
region surrounding NGC 6334 is an outlier in the distribution of field
directions determined from optical polarimetry, just as the NGC 6334 cloud
is an outlier in the distribution of cloud field directions determined by
SPARO.  In both cases the field direction corresponding to NGC 6334 is
rotated away from the direction of the plane by a large angle ($>$ 60\degr ).  This finding is consistent with our suggestion that field
direction tends to be preserved during GMC formation.  Finally, we compare
the disorder in our magnetic field maps with the disorder seen in magnetic field maps
derived from MHD turbulence simulations.  We conclude from these comparisons that the magnetic
energy density in our clouds is comparable to the turbulent energy
density.

\end{abstract}





\section{Introduction}
It is generally believed that the ionization fraction in interstellar
clouds is sufficiently high for flux-freezing to apply, and that
interstellar magnetic fields are thus well coupled to the gas.  Furthermore,
the fields may be strong enough to be dynamically important, and thus may
play important roles in star formation.  These issues have been reviewed by
Crutcher (2004).  Furthermore, as reviewed by Mac Low \&\ Klessen (2004),
magnetohydrodynamic (MHD) turbulence is probably an important player in the
star formation process.

Submillimeter polarimetry is a technique for mapping interstellar magnetic
fields (Lazarian 2000) that is well suited for studying dense star forming
clouds.  In particular, this method appears to have an advantage over the
related technique of optical/near-IR polarimetry of background stars.
Specifically, it has been shown by Whittet et al.\ (2001) and Arce et al.
(1998) that grain populations residing in regions that are shielded from the
interstellar radiation field have very low polarizing efficiency at
optical/near-IR wavelengths. Moderate shielding ($A_{v} = 1-2$ mag) seems to be
sufficient to sharply reduce the efficiency. By contrast, when we study the
submillimeter emission from highly obscured regions ($A_{v}$ $\sim$\ 30 mag) in a
dense cloud, we find significant polarization (Crutcher et al.\ 2004). Thus,
submillimeter polarimetry is especially useful for such regions.

This discrepancy between optical/near-IR polarization efficiency and
submillimeter efficiency, for highly obscured regions, is explained by Cho
and Lazarian (2005) as an effect of grain size.  Under the assumption that
grains are aligned by the radiative torque mechanism (Dolginov 1972, Draine
\&\ Weingartner 1996, 1997), they show that in highly obscured regions only
the largest grains will be aligned. They argue that these large aligned grains
dominate the submillimeter emission but are relatively less important for
the optical/near-IR extinction.

Maps of submillimeter/far-IR polarization usually cover relatively small sky
areas; typically of order ten square arcminutes (Greaves et al. 2003,
Hildebrand 2002, Dotson et al.\ 2000). Because Giant Molecular Clouds (GMCs)
usually extend over much larger sky areas, approaching a square degree,
these small-scale maps have not been very useful for probing the large-scale, 
or global, magnetic field of a GMC.\ Not every
submillimeter polarization map is small on the sky, however: The ARCHEOPS
balloon-borne submillimeter polarimeter (Beno\^{i}t et al. 2004) mapped the
degree-scale submillimeter (850 \micron) polarization of large sections of the
Galactic disk.  Binning their data to a resolution of several degrees, they
obtain significant detections of polarization with magnetic field directions
generally running parallel to the Galactic plane.  Because the large-scale
Galactic magnetic field determined from optical polarimetry is also known to
run parallel to the plane (Mathewson \&\ Ford 1970), the ARCHEOPS
result shows that the Galactic field penetrates into the dense gas that dominates
the submillimeter emission. In contrast, the smaller scale
far-IR/submillimeter polarization maps of GMCs in the disk show a wide range
of inferred field directions and no preferred orientation with respect to
the Galactic plane (see Fig. 21 of Hildebrand 2002).

The South Pole polarimeter SPARO is optimized for submillimeter polarimetry
on angular scales larger than those typically accessible from ground-based
submillimeter dishes, though not as large as those studied by ARCHEOPS.\ With 
SPARO we can map fields over a significant fraction a GMC while still
resolving field structure within the cloud.  Here we present SPARO
polarization maps for four GMCs in the Galactic disk. Each map extends over
a sky area corresponding to several hundreds of square arcminutes.

\section{Observations and Results}

The observations were made at Amundsen-Scott South Pole
station, using the Viper \mbox{2-m} telescope (Peterson et al.\ 2000,
Kuo et al.\ 2004)
together with Northwestern University's submillimeter polarimeter, SPARO
(Dotson et al.\ 1998, Renbarger et al.\ 2004).  
Radiation entering SPARO passes through a rotatable
half-wave plate, and is then divided into two orthogonal polarization
components, each of which is detected by a separate $^3$He-cooled 9-pixel
detector array.  SPARO thus measures polarization simultaneously at nine
sky positions, arranged in a 3 by 3 square pattern.  Because the South
Pole is an exceptionally good submillimeter site Lane (1998), SPARO
obtains extremely good sensitivity to spatially extended, low
surface-brightness emission. SPARO's first observations, made during
Austral Winter 2000, were reported by Novak et al.\ (2003).  Here we present
results from our second observing campaign that took place during
April--August 2003.

For the 2003 observations, the beam FWHM was determined to be 4\arcmin\
$\pm$ 0.5\arcmin, and the pixel-to-pixel separation was 3.3\arcmin.  The
pointing accuracy was $\pm$ 1\arcmin.  SPARO's spectral passband is
centered at $\lambda_0 = $ 450 \micron, with fractional bandwidth
$\Delta\lambda/\lambda_0=$ 0.10.  Our data acquisition scheme involves
carrying out standard ``photometric integrations'' at each of six
half-wave plate rotation angles, successively (Hildebrand et al.\ 2000).  Each
photometric integration, in turn, involves rapidly switching the array
footprint back and forth on the sky between the source position and a
reference position.  Reference signals are subtracted from source signals.  
We used two reference positions (Hildebrand et al.\ 2000), separated from the source
position by +0.65\degr\ and $-$0.65\degr\ in cross-elevation,
respectively.

Our data reduction procedures follow closely those described by
Hildebrand et al.\ (2000) with one important difference.  Hildebrand et al.\ (2000) describes how
the ``polarization signal'' is computed from each photometric integration
(see above) by taking the difference between signals from corresponding
pixels of the two detector arrays, and then dividing this by the sum of
these signals (see their equation 7).  This corresponds to dividing
polarized flux by total flux.  We found that for our observations the
total flux was often not well determined by a single photometric
integration, due to variations in atmospheric emission significantly above
the level of the photon noise.  These variations are usually referred to
as ``sky noise'', and they affect the total flux measurement but not the
difference signals.  In order to circumvent the problem, we developed a
new method for computing the polarization signal, which we now describe.

We grouped our data into ``sets'' consisting of nine identical ``half-wave
plate cycles''.  As described above, each such cycle involved six
photometric integrations, each taken at a different half-wave plate angle.  
For each set, we computed a single value for the total flux, or signal
sum, for each of the nine pixels.  We also computed, for each pixel, six
values of the difference signal, one for each half-wave plate position.  
These six difference signals were then divided by the set-average total
flux for the corresponding pixel.  Our new technique for determining the
polarization signal requires stable atmospheric transmission over time
scales one hour, corresponding to the duration of one set.  On the basis
of opacity measurements obtained with the CMU/NRAO tipper (Peterson
et al.\ 2003),
we estimate that signals were stable to $\pm$ 10\%.  A more detailed
description of the new data analysis procedure is given by Li et al.\ (2005),
and a similar procedure that was applied to data from the Hertz
polarimeter is described by Kirby et al.\ (2005).

The instrumental polarization was determined by calibrating on the Moon
and on the intensity peak of Sgr B2, following the same procedure that we
used for our 2000 observations (Novak et al.\ 2003, Chuss 2002).  The level of
systematic error in our measurements is $<$ 0.3\%, which translates into
an uncertainty in polarization angle of $< (9\degr)(1.0\%/P)$.

We observed four GMCs: NGC 6334, the Carina Nebula, G333.6-0.2, and G331.5-0.1. 
Three of our targets were chosen for their high column density and large angular extent as determined from the dust opacity map of Schlegel et al.\ (1998). This map is derived by combining COBE/DIRBE and IRAS/ISSA maps. One of our targets, Carina, has somewhat lower dust opacity than the others. It was chosen for its high elevation 
($\sim 60\degr $ at S.\ Pole) which corresponds to higher atmospheric transmission. 
In Tables 1 -- 4, we give the measured degree and angle of polarization and associated statistical errors for all 
sky positions having P $> 2\sigma _{P}$. Note that $\sim 80\%$ of the measurements have P $> 3\sigma _{P}$. 
All measured degrees of polarization are larger than 0.3\%, i.e.\ they are above 
the level of systematic error.   Figure 1 shows the inferred magnetic field directions (that are orthogonal to the measured polarization directions) and the magnitudes of polarization (denoted by the lengths of the bars) superposed on IRAS 100 \micron\ maps.

\section{Discussion}

In this section, we first give an overview of the characteristics of the four GMCs we  observed (\S\ 3.1). Then we  discuss the measured degrees of polarization (\S\ 3.2).  In the remaining four sections we deal with the inferred magnetic field directions, beginning with an overview of the statistics of these directions (\S\ 3.3), next comparing with mid-IR intensity maps (\S\ 3.4) and with information obtained from optical polarimetry of stars (\S\ 3.5), and finally discussing the relevance to theoretical ideas concerning formation and structure of GMCs (\S\ 3.6). 
\subsection{Characteristics of the observed clouds} \label{bozomath}
\subsubsection{The Carina Nebula}
Perhaps the best studied of our four targets is the Carina Nebula (NGC 3372), a very bright emission nebula excited by OB star clusters and lying at a distance of about 2.7 kpc (\S\ 3.1.5) in the Carina spiral arm.  Optical nebulosity and mid-IR emission extend over $\sim 3\degr$\ in Galactic latitude and $\sim 2\degr$\ in longitude (Smith et al. 2000), but most of the far-IR and radio continuum flux is concentrated in the central $1\degr\ x 1\degr$\ region shown in Figure 1 (Shaver \& Goss 1970; Zhang et al.\ 2001).  The associated molecular clouds are referred to collectively as the Carina Molecular Cloud Complex (Zhang et al.\ 2001).  This complex is elongated in a direction roughly parallel to the Galactic plane, with an extent of about three degrees ($\thicksim$ 140 pc) in Galactic longitude. The complex has a mass of about 7 x $10^{5}$ $M_{\odot}$ (Grabelsky et al.\ 1988).

The bright central radio peak of the Carina Nebula consists of two main
parts, Carina I and Carina II (Davidson \&\ Humphreys 1997; Whiteoak 1994;  
Brooks, Storey, \&\ Whiteoak 2001), whose sky locations are indicated in
Figure 1 with filled and open star symbols, respectively.  Carina I
(a.k.a.\ G287.4-0.6) is powered by the OB star cluster Tr 14, located a few
arcminutes to the East of Carina I.\  Carina II (a.k.a.\ G287.6-0.6) is
powered by another OB cluster, Tr 16, that contains Eta Carinae, a very
massive ($>$ 100 $M_{\odot}$) evolved star.  Eta Carinae is visible in
Figure 1 as the flux peak located a few arcminutes SouthEast of Carina II,
and this unusual star can also be seen in Figure 5, as the dominant source
near the lower left corner of the Carina map.

Tr 14 and Tr 16 contain large numbers of unusually massive stars.  Six of the 17 known O3-type stars in our Galaxy are found in these two clusters (Davidson \&\ Humphreys 1997).  Tr 14, which powers the region we mapped polarimetrically, is estimated to be only about one million years old (Walborn 1995).  There is little evidence for ongoing formation of massive stars within the region encompassed by Tr 14, Tr 16, Carina I, and Carina II (Davidson \&\ Humphreys 1997), but evidence for ongoing star formation has been found surrounding this region, especially to the SouthEast of Eta Carinae (Megeath et al. 1996, Smith et al.\ 2000).  For this reason, Carina is often cited as an example of sequential star formation (de Graauw et al. 1981, Smith et al.\ 2000).

\subsubsection{G333.6-0.2 and NGC 6334}
The H\footnotesize II\ \normalsize region G333.6-0.2, although unimpressive at optical wavelengths, is one of the brightest radio sources in the Southern sky (Goss \&\ Shaver 1970).  When its infrared counterpart was discovered by Becklin et al. (1973), they noted that its luminosity in the 1 - 25 \micron\ range was unsurpassed by any other H\footnotesize II\ \normalsize region.  G333.6-0.2 is the bright peak at (0\degr, 0\degr) in Figure 1.  McGee, Newton, \&\ Butler (1979) mapped many other radio sources near this peak. These are often collectively referred to as ``the $l = 333\degr$\ complex'' (de Graauw et al. 1981).  At least four of these other sources can be seen in Figure 1, but only about half of the emission from the $l = 333\degr$\ complex falls within the boundaries of Figure 1, as the complex extends beyond the Southern and Western edges of the image.  Maps of the associated CO emission have been obtained by de Graauw et al.\ (1981) and Bronfman et al.\ (1989).  The radial velocity of the $l = 333\degr$\ complex is about -50 km/s and the distance is about 3.0 kpc (Sollins \&\ Megeath 2004; \S\ 3.1.5).  The complex is elongated parallel to the Galactic plane (McGee et al.\ 1979; Russeil et al.\ 2005; Fig.\ 1) and has an extent of $\sim$\ 1.5\degr\ in longitude (estimated from either radio or molecular maps) corresponding to 80 pc.  Cheung et al.\ (1980) estimate a mass of $10^{5}$ $M_{\odot}$ for a region $\thicksim$ 4 pc in extent centered on G333.6-0.2. The total mass of the complex must be much larger, perhaps comparable to that of the Carina Molecular Cloud Complex (\S\ 3.1.1).  

NGC 6334 is an optically visible H\footnotesize II\ \normalsize region lying at a distance of 1.7 kpc (Neckel 1978; \S\ 3.2). The radio map of Goss \&\ Shaver (1970) shows ionized gas distributed over a region $\thicksim$ 10 pc in size, while the optical nebulosity extends over a region about 20 pc in extent (Gardner \&\ Whiteoak 1975; Straw \&\ Hyland 1989). The extent of the associated molecular gas is also about 20 pc (Dickel, Dickel, \&\ Wilson 1977). From these maps we see that gas associated with NGC 6334 may not extend much beyond the region shown in Figure 1. Based on the CO map of Dickel et al.\ (1977), Straw \&\ Hyland (1989) estimate a mass of 1.5 x $10^{5}$ $M_{\odot}$ for the entire cloud. Comparing the size and mass of NGC 6334 with values given above for the Carina molecular cloud complex and the $l = 333\degr$\ complex, we see that NGC 6334 may be somewhat smaller and less massive.  

Sollins and Megeath (2004) studied both G333.6-0.2 and NGC 6334. For both sources they present strong evidence of ongoing formation of massive stars. In NGC 6334, they studied the source NGC 6334 I, that lies toward the Northern edge of the dense ridge of emission seen in Figure 1.  For both G333.6-0.2 and NGC 6334 I, they find molecular cores containing $\thicksim$ $10^{5}$ $M_{\odot}$ of dense ($\ga$ $10^{6}$ $cm^{-3}$) gas, as well as young H\footnotesize II\ \normalsize regions. Specifically, G333.6-0.2 corresponds to a compact H\footnotesize II\ \normalsize region and NGC 6334 I to an ultra-compact H\footnotesize II\ \normalsize region. 

Based on their distances (\S\ 3.1.5) and on the Galactic spiral arm 
model of Taylor \&\ Cordes (1993) NGC 6334 and G333.6-0.2 are located in the Carina-Sagittarius and Scutum-Crux spiral arms, respectively.

\subsubsection{G331.5-0.1 }
Of our four targets, G331.5-0.1 is the one that has been studied least.  This H\footnotesize II\ \normalsize region and six other nearby radio sources form a group having an extent of about a degree in Galactic longitude, and about 0.5\degr\ in latitude (Amaral \&\ Abraham 1991).  The group is centered near $l = 331\degr$. Several of these radio sources can be seen in Figure 1, and several others lie beyond the beyond the Southern and Western edges of the image.  

Not all the members of this group have the same line-of-sight velocity, so it is generally believed that the group is a superposition of two unrelated complexes (Russeil et al. 2005, Amaral \&\ Abraham 1991, Caswell \&\ Haynes 1987). Both complexes are elongated parallel to the Galactic plane, and they are separated not only in velocity but also in Galactic latitude.  The complex containing G331.5-0.1, which is seen at (0\degr, 0\degr) in Figure 1, has generally more positive latitudes and a higher negative velocity.  This complex is referred to by Russeil et al. (2005) as ``the -87 km/s complex''.  The other complex, which contains the source G331.3-0.3, seen at (0.0\degr, -0.3\degr) in Figure 1, has generally more negative latitude and a lower negative velocity.  It is referred to by Russeil et al.\ (2005) as the -65 km/s complex. Except for G331.3-0.3, all major flux peaks seen in Figure 1 correspond to the -87 km/s complex.

For the most part, sky locations where we obtained polarimetry data correspond to the  -87 km/s complex.   About seven vectors lying at the SouthEast corner of our polarization map are closer to G331.3-0.3 than to any source in the -87 km/s complex, so they probably correspond to the -65 km/s complex.  Note that many of these vectors are more nearly perpendicular to the Galactic plane, while the vectors for the rest of the map are more nearly parallel to it.

The distance to the -87 km/s complex is still uncertain, but Russeil et al.\ (2005) argue for a distance of 5.3 kpc (\S\ 3.1.5).  The -65 km/s complex lies at 4.2 kpc, according to these authors.  Molecular emission corresponding to the -87 km/s complex can be seen in the CO maps of Bronfman et al.\ (1989). The size and shape of this CO structure is similar to that measured for the \textit{\textbf{l}} = 333\degr\ complex that contains G333.6-0.2 (\S\ 3.1.2) but the integrated intensity is somewhat lower.  Factoring in the greater distance, the molecular mass of the -87 km/s complex could be comparable to that of the \textit{\textbf{l}} = 333\degr\ complex. Despite the overall paucity of observations for G331.5-0.1, it does seem that evidence for ongoing formation of massive stars has been found.  For example, at the position of this source we find a massive ammonia core (Vilas-Boas \&\ Abraham 2000), CS emission (Bronfman, Nyman, \&\ May 1996), and strong methanol maser emission (Walsh et al. 1997).

\subsubsection{Comparison of the four GMCs observed }

All four SPARO targets are associated with massive GMCs ($10^{5}$ - $10^{6}$ $M_{\odot}$) and ongoing or recent formation of massive stars.  Using IRAS results as their database, Kuiper et al.\ (1987) did a comparative study of 65 of the brightest southern molecular clouds in the Galaxy, including our four targets.  They calculated 60/100 \micron\  color temperatures, and column densities.  For NGC 6334, G333.6-0.2, and G331.5-0.1, temperatures were in the range 40-43 K.  These values are near the peak of the distribution for the 65 sources.  For the Carina Nebula, a color temperature of 48 K was found.  This value lies in the upper tail of the distribution.  Column densities for Carina were significantly lower than for our other three targets.  One explanation for these differences is that Carina is more evolved than the others, so more gas has been dispersed and more energy has been released to heat the dust.  This would be consistent with the observed fact that massive stars do not seem to be presently forming within the region corresponding to our polarimetric map of Carina (\S\ 3.1.1).

\subsubsection{Methods used to determine cloud distances}
For purposes of comparing our results with optical polarimetry (\S\ 3.5) and estimating physical scales corresponding to the SPARO beam size (\S\ 3.6) we require accurate estimates of the distances to our targets.
Distance determinations for these targets fall into two categories: stellar distance estimates (e.g.\ spectroscopic parallax) and kinematic distance estimates, based on the Galactic rotation curve.  The latter solution is multi-valued, providing a ``near distance'' and a ``far distance''.  Using the spectroscopic parallax method, Neckel (1978) obtained distance estimates for OB stars presumed to be associated with NGC 6334.  On this basis, they derive a distance of ($1.7 \pm 0.3$) kpc for this cloud.

The star clusters Tr 16 and Tr 14 that power the Carina Nebula are separated by about 15 pc on the sky.  The maps in Smith et al.\ (2000) show that this nebula is reasonably isolated on the sky, so it seems likely that the two clusters lie at roughly the same distance.  Distances to these clusters and to specific stars within them have been obtained by Carraro et al.\ (2004), Davidson et al.\ (2001), Freyhammer et al.\ (2001), Rauw et al.\ (2001), Tapia et al.\ (2003), and Vazquez et al.\ (1996).  These six independent estimates have a mean of 2.7 kpc and a standard deviation of 0.4 kpc.  Our distance estimate does not change significantly if we drop the requirement that the two clusters must lie at the same distance and instead simply estimate the distance to Tr 14, which is the cluster that powers the far-IR flux peak that we mapped polarimetrically with SPARO.

For G333.6-0.2 and G331.5-0.1, distance estimates are kinematic. For G333.6-0.2, we adopt the value 3.0 kpc, given by Sollins \&\ Megeath (2004) and by Colgan et al. (1993). Other values found in the recent literature are 3.5 kpc (Russeil et al. 2005) and 2.8 kpc (Vilas-Boas \&\ Abraham 2000). The far kinematic distance of about 11 kpc can be ruled out because Russeil et al. (2005) detect optical emission from six sources in the $l = 333\degr$\ complex (which they refer to as ``CO cloud I''). For G331.5-0.1, it is again the near distance that is almost always 
quoted, but in this case it is not as easy to rule out the far distance. The strongest discriminant seems to be the comparison of absorption and emission spectra carried out by Dickey et al.\ (2003) that places G331.5-0.1 at the near kinematic distance, which is given by Russeil et al.\ (2005) as 5.3 kpc.

\subsection{Measured polarization magnitudes}
Hildebrand et al.\ (1999) show the distrubution of measured magnitudes of 350 \micron\ polarization for a large
sample of measurements, obtained from
polarization maps that generally sample smaller spatial scales in comparison with
our SPARO maps. The mean magnitude in the Hildebrand et al.\ (1999) sample is 1.6\%, while ours is slightly higher 
at 2.0\%. Several factors could account for this difference. Firse note that in Figure 1, the higher degrees of polarization usually are found in regions with lower column density, a trend that has also been seen in smaller-scale maps (Matthews et al., 2000; Henning et al., 2001; Lai et al., 2002; Crutcher et al., 2004). Since the SPARO data sample regions of generally lower column density in comparison with the regions studied by Hildbrand et al.\ (1999), this effect could explain why SPARO sees higher polarization. On the other hand, SPARO's large beam averages over more field disorder, which would tend to reduce the polarization magnitude. Another effect that could be important is our improved data analysis method (\S\ 2; Li et al. 2005), that removes the bias toward low polarization magnitude. Finally, note that Vaillancourt (2002) present
evidence for structure in the polarization spectrum, so the modest
wavelength difference (350 \micron\ vs.\ 450 \micron ) could be important.

\subsection{Statistics of the inferred magnetic field directions}
Figure 2 shows the distribution of projected magnetic field position
angles, for the entire sample of measurements for all four clouds.  The
position angle is given in Galactic coordinates, with 0\degr\
corresponding to Galactic North-South, and with position angle increasing
in the counter-clockwise direction.  Note that the measurements tend to
cluster near the dark vertical line at position angle 90\degr ,
corresponding to magnetic fields running parallel to the Galactic plane.  
Relatively few measurements are found near the left- and right-hand
extremes of the histogram, corresponding to fields orthogonal to the
plane.  We can explain this by supposing that large-scale magnetic fields
in GMCs are generally parallel to the even larger scale Galactic fields in
the regions surrounding these clouds, which are known to run
preferentially parallel to the Galactic plane (Mathewson \&\ Ford 1970).  
Much of the remainder of this paper is devoted to exploring the validity
of this conclusion, and its implications.

As a first step, we break the histogram in Figure 2 into four separate
histograms, one for each of our four target GMCs.  This is shown in Figure
3, where we also show vertical dotted lines indicating the mean position
angle for each cloud.  The method we use to calcuate this mean is
described in the paragraphs below.  For three of the four clouds, the mean
field direction is nearly parallel to the Galactic plane, but for NGC 6334
the mean is almost perpendicular to the plane.

Because the field direction measurements wrap around, with 0\degr\ =
180\degr , the mean position angle is actually undefined.  However, it is
clear that each cloud has a well defined peak in the distribution of
angles (Fig.\ 3).  We have devised a technique for estimating the position
of this peak, based on the formalism of Stokes' parameters (e.g., see
Jackson 1999).  Two of the four Stokes' parameters, Q and U, contain
information about the state of linear polarization of a light beam.  They
are related to the intensity of linearly polarized flux $I_{PF}$ and angle of
polarization $\phi$ via
\begin{eqnarray}
 Q = I_{PF}cos(2\phi)\ ;\ U = I_{PF}sin(2\phi). 	
\end{eqnarray}

Our method is as follows:  For a given cloud, we first compute $Q_{i}$ and $U_{i}$
for each sky position i where we have detected polarization.  Note that by
summing these to form $Q_{sum}$ and $U_{sum}$, one can determine the polarization
state that would be measured by an imaginary polarimeter that acts by
combining all flux from the region studied by SPARO and making one
polarization measurement on it. However, since our goal is to study the
global field of each cloud, we do not want regions having high column
density to be given more weight than regions having low column density.  
Furthermore, we do not want regions having high grain alignment efficiency
to be given more weight than regions with low efficiency.  Accordingly, to
ensure that each sky position will be given equal weight, we compute
\begin{eqnarray}
 Q'_{i} \equiv\ Q_{i} / I_{PF}\  ;\  U'_{i} \equiv\ U_{i} / I_{PF},  	
\end{eqnarray}
and then average all values of $Q'_{i}$ and $U'_{i}$ for a given cloud to
form $\bar{Q'}$ and $\bar{U'}$. From these ``equal weight'' average values
we can derive a mean polarization angle for the cloud using the formalism
in equation (1). We refer to the mean magnetic field direction obtained
in this way as the ``equal weight Stokes mean'', $\theta_{ewsm}$. Note
that because the total flux and degree of polarization cancel out in
equation (2), $\theta_{ewsm}$ is actually determined using only the
measured angles of polarization. No other information is required.
Our method for computing $\theta_{ewsm}$ also yields another parameter
that we refer to as the ``order parameter'', defined as 
o.p. = $\sqrt{(\bar{Q'})^2 + (\bar{U'})^2}$. For a set of polarization
measurements all having the same position angle, we obtain o.p. = 1.  For
a set of measurements having position angles uniformly spaced over a full
180\degr, we obtain o.p. = 0.  We will use this parameter in our analysis
of optical polarimetry data (\S\ 3.5).

Note that for three of our four clouds, the mean field angle
$\theta_{ewsm}$ is within 15\degr\ of the Galactic plane (Figs.\ 3 \&\ 4).  
For a set of clouds having a random distribution of $\theta_{ewsm}$ we
would expect on average only one cloud in six to have $\theta_{ewsm}$
within 15\degr\ of the plane.  If we choose four random angles between
0\degr\ and 180\degr, the probablility for obtaining at least three of the
four within the interval $90\degr\ \pm\ 15\degr$ is only $(4 \times\ 5 +
1)\ /\ 64 = 0.0162$. Therefore this is unlikely to be a chance alignment.  
The most likely explanation is the one we advanced in the first paragraph
of this section: Large-scale magnetic fields in GMCs are preferentially
parallel to the fields in the even larger scale regions of the Galaxy that
surround these clouds.  We explore this issue further in \S\ 3.5, where we
also examine the case of the discrepant cloud NGC 6334.    

\subsection{Comparisons with MSX maps}
Polycyclic aromatic hydrocarbon molecules (PAHs) can be used as a tracer
of massive star formation (Peeters 2004). These molecules absorb far-UV photons emitted by stars and transfer the energy into vibrational modes associated with molecular stretching and bending. The PAHs then cool by emitting IR radiation, mostly in the wavelength
range 3 -- 11 \micron. The 8 \micron\ emission shown in Figure 5 is mostly, though not entirely, due to PAHs. The main exception is that some of the point sources seen in these images corresepond to stars surrounded by hot dust.    

The 3 -- 11 \micron\ PAH spectrum depends significantly on the charge states (see Fig.\ 2 of Allamandola et al.\ 1999; Peeters 2002), which are determined by the ratio of the illuminating UV radiation field and the electron density, $G_{0}/n_{e}$. In regions having high $G_{0}/n_{e}$ ratio the ionized states of PAH dominate, while for low $G_{0}/n_{e}$ ratio it is the neutral states that are more common. In the ionized state, the intensity of the 8 \micron\ emission is much higher. In crossing the boundary of an H\footnotesize II\ \normalsize region, $n_{e}$ changes dramatically but $G_{o}$ changes relatively little. The result is a sharp change in the 8 \micron\ brightness at the boundary of an H\footnotesize II\ \normalsize. This effect can be seen in the 8 \micron\ map of Carina shown in Figure 5.  The
shapes of the bipolar H\footnotesize II\ \normalsize bubbles seen in this figure are very close to shapes seen in the superbubble simulations of Silich \&\ Franco (1999).  
     
     OB stars ionize their surroundings, and the ionization pressure wave propagates outwards from the star-forming region. The speed of propagation is inversely proportional to the density of the ambient interstellar medium, so H\footnotesize II\ \normalsize regions usually grow fastest along the direction perpendicular to the Galactic plane.  This explains the
bipolar morphology of the H\footnotesize II\ \normalsize bubbles (Silich \&\ Franco 1999, Smith et al. 2000).

  Note that the magnetic field directions we determined for Carina closely follow the boundaries of the two expanding H\footnotesize II\ \normalsize bubbles (Fig.\ 5).  This can be understood as the effect of the expansion of the bubble upon the magnetic field. In this model, we must assume that the expansion of the ionization front is accompanied by bulk gas motion, such that gas is generally flowing in a direction that points away from the center of the bubble. This is reasonable as the ionized gas will be heated and will tend to expand.  In this case the gas just beyond the edge of
each bubble will be compressed in a direction perpendicular to the bubble edge. Under flux-freezing conditions, such compression will have a strong effect on the magnetic field. The result will be that the field will tend to be parallel to the compression front (e.g., see Novak et al.\ 2000), precisely as we observe in Carina. 

However, there is an alternative explanation that could account for the observed parallelism between the edges of the bubble and the field without requiring a compression front. To see this, note that gas expansion, under flux-freezing conditions, will always cause distortions in the ambient magnetic field regardless of the existence or non-existence of a compression front external to the bubble. A concrete example is provided by the model of Tomisaka (1998) where it is the magnetic tension rather than external gas pressure that acts to resist the expansion of the bubble. 

The bubble-like structures seen in Carina are much more striking than the structures seen in the three other 8 \micron\ maps shown in Figure 5. This may indicate that, in comparison with our other three targets, Carina has suffered a relatively greater amount of disruption induced by star formation.  If so, that would explain why Carina shows more variation in inferred magnetic field direction in comparison with the other SPARO targets (Fig.\ 3).  This interpretation is consistent with what we learned from the comparison between the four clouds (\S\ 3.1.4).  

High-resolution radio polarization measurements of galaxies show that the magnetic field is well aligned with spiral arms (Neininger 1992, Neininger \&\ Horellou 1996, Patrickeyev et al.\ 2005). Because field disorder is magnified when observed along the field direction, one might argue that the lesser degree of uniformity of the fields in Carina is a result of the near-coincidence of the line-of-sight and the tangent to the Carina spiral arm at the location of the cloud (The angle between line-of-sight and the arm tangent is $\sim$\ 20\degr, much lower than for NGC 6334 or G333.6-0.2 ).  However, the strong correlation between the fields and the bubble boundaries shows that the main reason for the field disorder is star formation. 

G333.6-0.2 also contains bipolar PAH bubbles.  These can be seen in the image shown in Figure 5, at position (0.10\degr, -0.05\degr) and (0.0\degr, 0.05\degr). They are much easier to see in the recently obtained Spitzer/GLIMPSE images (B.\ Whitney, private communication). In contrast to the case of Carina, our polarization map for G333.6-0.2 covers a region significantly larger than the area containing the bubbles. The field in
G333.6-0.2 is basically uniform, but note that the field lines running near the ``Galactic North'' edge of the northern bubble appear to be pushed outwards by the bubble. NGC 6334 also shows bubble-like structures in Figure 5, but the strongest such structures occur at locations for which we have no SPARO polarization data. 

\subsection{Comparison with stellar polarization data}
It has been shown that Galactic magnetic fields tend to be parallel to the Galactic disk on large scales. For example, optical polarization measurements for stars more distant than
2 kpc show inferred field directions that are mostly parallel to the plane (Mathewson \& Ford 1970). As discussed in \S\ 3.3, our observations suggest that on the scales of our SPARO polarization maps (much smaller than 1 kpc), this tendency is still evident. This can be understood if the processes that act to accumulate diffuse gas and thereby form a GMC do not significantly alter the mean magnetic field direction in the gas. But in this case, how do we account for the outlier NGC 6334? One answer to this question is suggested by the fact that even though the large-scale Galactic field is parallel to the disk, there are nevertheless spatial fluctuations in the Galactic field that occur on scales that are larger than the size of a GMC (e.g., Mathewson \&\ Ford 1970). If NGC 6334 happened to form in a region where such fluctuations had resulted in a field significantly different from the large-scale
average, then this could explain the discrepancy. 

In principle, this idea can be tested by comparing SPARO results for each cloud with the field direction in the surrounding diffuse medium as sampled by optical polarimetry of stars.  In practice, we have only been able to do this for NGC 6334, the closest of our four targets.  Here we describe the results of this analysis, that was carried out using a stellar polarimetry database published by Heiles (2000). Since this database was created by combining many previous polarization catalogs, it is referred to by the author as an ``agglomeration''. When computing the field direction in the surrounding diffuse medium, we must choose a length scale to characterize this larger region. The natural choice is the accumulation length, defined so that the cube of this length corresponds to the volume of diffuse ISM containing a gas mass equal to 
the mass of a GMC.  Williams et al.\ (2000) estimate an accumulation length 
of 400 pc for the Galactic disk.    
    
    A stellar polarization measurement gives the mean field direction, weighted by density of dust particles, along the line of sight. To determine the field in a particular region, one must subtract the foreground polarization, inferred from foreground stars, from that measured for background stars (e.g., Marraco et al.\ 1993). In our analysis of the Heiles (2000) database, we carry out such subtractions. Stellar distance measurements are important because we will rely on them to define the foreground and background stars for a given region. Based on comparisons among the various catalogs that he has agglomerated, Heiles (2000) estimates a typical distance uncertainty of about 20\%.

\subsubsection{Evidence of Malmquist bias in the Heiles database}
Besides stellar distance, another factor that is relevant to the problem of foreground-effect subtraction is extinction. For the Heiles (2000) database, Figure 6 shows how the selective extinction, E(B-V), varies with distance. Each point represents the mean E(B-V) for stars within a 100 pc distance interval. In making this figure, we have included only stars having $\left|b\right|\ \leq\  0.1\degr$, where $b$ is the Galactic latitude. The figure shows that the selective extinction grows quite linearly up to $\sim\ 1.5$ kpc, with a slope close to the 0.6 mag/kpc value given by Spitzer (1978). Beyond this distance, the extinction tends to level off. 
There are three effects that could contribute to this flattening of  the extinction curve: (1) There is more dust nearer to the Sun than further from it (i.e., we live within a local density enhancement). (2) The data base contains Malmquist bias. This refers to the fact that in a flux-limited sample, observers preferentially see atypically bright objects at larger distances. Two factors that affect brightness are luminosity and extinction.  From the point of view of this discussion, the latter possibility is interesting, because if there is a bias toward
lower extinction at large distance then this can explain
the flattening of the extinction curve. (3) For a fixed latitude, sight-lines to more distant stars extend to higher distances above Galactic plane, where there is less dust.

Because we have restricted b to $\pm\ 0.1\degr$, we can be reasonably sure that explanation (3)  does not play a large role.  This is because no sight line in our sample extends further than 10 pc from the Galactic plane, for distances up to 3 kpc. 10 pc is much smaller than the scale height of the H\footnotesize I\ \normalsize gas (Spitzer 1978). Given that the dust density should be highest at the location of the spiral arms, and that the Sun is situated about half-way between two such arms, each about 2 kpc distant (Xu et al.\ 2006), it seems highly unlikely that explanation (1) could account for all of the flattening seen in Figure 6. Thus, explantion (2) must play a significant role.

Malmquist bias can introduce uncertainties in the correction for foreground polarization. For this reason, we believe that using the data of Heiles (2000) to study regions much beyond 1.5 kpc will give unreliable results. We have chosen to study regions as far as 1.7 kpc so that we can include the vicinity of NGC 6334 (d = 1.7 kpc; \S\ 3.1.5).  The next closest source, Carina, lies at a distance of 2.7 kpc, well into the flat part of the extinction curve.  In \S\ 3.5.4, we show that Malmquist bias does not affect our main conclusions for regions as far as 1.7 kpc. 

\subsubsection{Choice of cells for analysis}
In our analysis of the Heiles (2000) database, we divide the nearby ($d
\leq\ 1.7$ kpc) regions of the Galactic plane into ``cells'' with
dimensions roughly corresponding to the accumulation length ($\sim\ 400$
pc; second paragraph of \S\ 3.5). One such cell is centered on NGC 6334.  
For each cell, we collect data from Heiles (2000) and apply a correction
for foreground polarization in order to estimate the mean magnetic field 
direction within each cell. For the
NGC 6334 cell, we compare this result with the mean field direction
inferred from the SPARO data. The other cells serve as comparison regions.

Our cells are cuboids in ($l$, $b$, $d$)-space, where $l$ and $b$ are
Galactic longitude and latitude and $d$ is distance from the Sun.  All
cells are centered on $b = 0\degr$, and centered on one of three possible
values of $d$: 0.7 kpc, 1.2 kpc, and 1.7 kpc.  As viewed from far above
the Galactic plane, they tile the local region of the Galactic disk,
forming three complete 360\degr\ rings of cells, with the whole pattern
centered on the Sun.  The angular size of a cell along its $l$-dimension
decreases with distance so that we can keep the corresponding spatial
scale (measured in pc) is approximately constant.  Thus, the number of
cells in a given ring increases with the $d$-value of that ring.  The
cells' angular size along the $b$-dimension also decreases with distance,
for the same reason.  The cells' linear sizes deviate somewhat from 400
pc, especially along the $b$ dimension, for reasons explained below.

For each cell, we construct foreground (background) cells that are similar
to the main cell but centered on the near (far) face of the main cell (see
Fig.\ 7).  As described in the next section, the polarization measurements
for ``foreground stars'' (stars lying within the foreground cell) are used
to correct the polarization measurements for each ``backgound star''
(star lying in the background cell), thus providing a set of rough
estimates for the stellar polarization induced by dust lying within the
main cell.

As noted above, our intent was to have cells with dimensions corresponding
to the accumulation scale, estimated to be 400 pc.  We modified the
in-plane spatial footprint of each cell, ($\delta l, \delta d$), from (400
pc, 400 pc) to (300 pc, 500 pc) in order to take into account the
uncertainty in stellar distances. This is why the three rings are
separated by 500 pc in $d$.  At the distance of our furthest background
cells, centered at $d$ = (1.7 + 0.25 = 1.95) kpc, the 20\% distance
uncertainty translates to approximately 400 pc. Thus, even after
increasing $\delta d$ from 400 to 500 pc, distance uncertainty is still an
important limitation to the accuracy of our analysis. However, increasing
$\delta d$ to 600 pc or higher while preserving non-overlapping cells
would have resulted in a sample containing only two rings of cells rather
than three, as there would not have been enough space for the foreground
cells corresponding to the closest ring of main cells.  

For each ring, we adjusted $\delta l$ slightly from 300 pc so that the
full 360\degr\ are covered by an integer number of cells. For the 1.7 kpc
ring, the longitudinal dividing lines of the cells are chosen so that one
cell is centered on NGC 6334. For the two closer rings, we have
arbitrarily chosen $l = 0\degr$ as a dividing line.

The size of the cells along the direction perpendicular to the Galactic
plane ($\delta b$) has been set at 120 pc. The reason that we chose such a
relatively small value for $\delta b$ is related to the scale height of
Galactic gas. Since the thickness of the H\footnotesize I\ \normalsize disk is 240 pc (Spitzer
1978), setting $\delta b$ to 400 pc would have resulted in the inclusion
of relatively high-latitude background stars having relatively little dust
within the main cell. The polarization induced by this relatively small
column of dust would have been very small, so the inferred
foreground-corrected polarization angles corresponding to these stars
would have been very uncertain. Because our method for determining the
mean magnetic field angle for each cell (\S\ 3.5.3) gives equal weight to
each background star, the use of $\delta b$ = 400 pc would have thus
resulted in loss of accuracy. We chose $\delta b$ to equal half of the
thickness of the HI disk in order to preferentially sample the denser part
of this gas layer. Note also that all four of the GMCs observed with SPARO
lie within 30 pc of the Galactic plane.  The resulting $\delta l \times\
\delta b$ sizes are $24\degr\ \times\ 10\degr$, $15\degr\ \times\
6\degr$, and $\ 10\degr\ \times\ 4\degr$ for rings at $d$ = 0.7, 1.2,
and 1.7\ kpc, respectively. Finally, note that for the NGC 6334 cell we
have explored the effect of increasing $\delta b$ (\S\ 3.5.5).

\subsubsection{Selection and processing of stellar polarization data for a single cell}
In order to correct for the effects of foreground polarization, we use the method introduced by Marraco et al.\ (1993). For each star in the background cell, we first collect a subset of the stars in the foreground cell, chosen to have sky coordinates reasonably close to those of the background star (Fig. 7). Then linear functions of $l$ and $b$, $q(l,b) = c_{1} + c_{2}l + c_{3}b$ and $u(l,b) = c_{4} + c_{5}l + c_{6}b$, are fit to the normalized Stokes parameters, $q_{f}$  and $u_{f}$, of the subset of foreground stars. Next, the contributions of the dust in the main cell to the Stokes parameters of the background star, $q_{b}$ and $u_{b}$, are estimated by $q_{b}-q(l_{o},b_{o})$ and $u_{b}-u(l_{o},b_{o})$, where ($l_{o},b_{o}$) are the coordinates of the background star. In this way we can estimate the polarization introduced to the background star by dust in the particular main cell under study. We will refer to this as the polarization residue, and it contains information about the magnetic field direction in the cell.

During the analysis, several criteria are used to reject stars. First, stars with nonpositive values for selective extinction are not used. Also, a background star is rejected if it has fewer than four corresponding foreground stars (Fig.\ 7). Although three stars are sufficient for the fit, we found that occasionally they will lie close to a straight line on the sky thus making the fit very unreliable. A background star having polarization residue below 0.2\% is also rejected as being too close to the 0.1\% accuracy of the Heiles (2000) database.

For three cells in our sample, we show in Figure 8 the polarization residues for all unrejected background stars. The mean field direction for a given cell is estimated by the equal-weight-Stokes mean $\theta _{ewsm}$ (\S\ 3.3) of the polarization residues. Cells containing fewer than five polarization residues are rejected. 22 of the 75 cells in our study survive this cut. 
A further cut is made based on the order parameter (o.p.; \S\ 3.3). Figure 8 illustrates how the order parameter is related to the degree of uniformity in the field directions given by the polarization residues. We see that for the cell with o.p. = 0.21, one would not conclude that the overall field direction for the cell is constrained in any way, so the mean field direction determined for this cell has little meaning. For the cell with o.p. = 0.41, however, the vectors suggest that there is a well-defined mean field direction. The criterion we used is o.p. $\geq$\ 0.3, which results in rejection of half of the remaining 22 cells.  

Using the analysis described above, we obtain mean field directions for 11 cells, and their distribution is shown in Figure 9. More than half of the cells have magnetic field angles within 30\degr\ of the Galactic plane, and the equal weight Stokes' mean of these 11 angles is 103\degr), which is reasonably close to the direction of the Galactic plane. The cell corresponding to NGC 6334 is shaded in Figure 9. Note that NGC 6334 falls in the same bin in this figure as it did in Figure 4 where we showed the distribution of the GMC field directions determined by SPARO. The polarization residues for the NGC 6334 cell are shown in the top panel of Figure 8.

\subsubsection{Correction for the effects of Malmquist bias}
As discussed in \S\ 3.5.1, there is evidence for Malmquist bias
in the Heiles (2000) database.  If we assume that this bias is the
only effect contributing to the flattening of the extinction curve of
Figure 6, then we can derive a ``bias correction'' that we can apply to the
Heiles database in order to remove the effects of the Malmquist bias.
Although it is not clear that all of the flattening is due to Malmquist bias (\S\ 3.5.1), we will nevertheless apply this
correction in order to get a rough idea of the possible effects of
the bias on our analysis.

Our bias correction is based on the reasonable assumption that the bias
affects the normalized Stokes parameters $q$ and $u$ in the same way that it affects
the measured values of selective extinction. The correction works as
follows: First we extend an extrapolation of the linear portion of the
extinction curve over the full range of distances (see dotted line in Fig.\  
6). We assume that this would be the measured extinction curve in the
absence of bias. Next, we multiply all of the normalized Stokes'
parameters measured for stars having $d \geq 1.5$ kpc by a correction factor
that depends on distance and is determined by taking the ratio of the
extinction value given by the extrapolation (dotted line) to
that given by the data themselves (estimated by the horizontal line).

If we apply this bias correction to the Heiles' data before using it in
our analysis, we find that 10 cells (rather than 11) survive the various
cuts. The resulting distribution of magnetic field directions for these
cells is shown in Figure 10. Note that the peak corresponding to fields
parallel to the Galactic plane is somewhat stronger. The NGC 6334 cell is
again indicated as a shaded box, and its field direction is essentially
unchanged. (Note that many of the stars we use in our analysis have
distances less than 1.5 kpc, so for these stars the bias correction has no
effect.)

There are two reasons to suspect that our correction for Malmquist bias
may be too strong. First, if Malmquist bias is not the only reason for
the flattening of the extinction curve, then we will be over-correcting.  
The second reason is that some of the cuts
we made on the data may have served to eliminate cells with
relatively stronger Malmquist bias. (This might be expected, as bias tends
to dilute the ``real'' polarization residues, due to actual grain alignment
by real magnetic fields, with spurious polarization effects due to bias.)  
This in turn might be expected to reduce the agreement among vectors in a
given cell and lead to rejection of the cell. In this case, applying the
correction to the surviving cells, that have relatively weaker Malmquist
bias, will result in over-correction. Keeping these possibilities in mind,
all we can conclude from our analysis is that our best estimate for the
true distribution of field directions probably lies somewhere between the
histograms of Figures 9 and 10.

In \S\ 3.3, we noted that three of our four GMCs have mean magnetic
field directions lying very close to the Galactic plane, and that the
probability of this being due to chance alignment is very low. There remained
the problem of understanding the outlier, NGC 6334.  The result of our
study of the local Galactic field on $\sim 400$ pc scales is that the NGC 6334 region turns
out to again be an outlier. Specifically, in terms of the angular
displacement of its field from the orientation of the Galactic plane, the
NGC 6334 cell is either the third highest of 11 (Fig.\ 9), or the highest
of ten (Fig.\ 10), depending on whether or not the bias correction is
applied.  We conclude that the most likely explanation for the discrepant
SPARO result for NGC 6334 is that the cloud formed in a somewhat unusual
region of the Galaxy, where the field direction (on $\sim\ 400$ pc scales) is
nearly perpendicular to the Galactic plane.  As a whole, our results for
the four clouds suggest that the mean field direction inside a GMC is
roughly parallel to that in a surrounding region having size approximately
given by the accumulation length.

\subsubsection{Dependence on assumed distance of NGC 6334 and cell size}
We concluded above that the cell corresponding to NGC 6334 is an outlier
with respect to the distribution of cell field directions. Next we show
that this conclusion is reasonably robust in the sense that it does not
depend sensitively on the precise distance assumed for NGC 6334, nor
on the precise cell dimensions.

As noted in \S\ 3.2, the uncertainty in the distance to NGC 6334 is
given as 0.3 kpc. This is due to systematic, not statistical errors
(Neckel 1978). We have repeated the analysis for this cell while
dispacing its center in 100 pc increments along the line-of-sight away
from the nominal value of 1.7 kpc. For distance values of 1.5, 1.6, 1.7,
1.8, and 1.9 kpc, we find magnetic field directions (measured counter-clockwise from
 Galactic North) of -21\degr, 5\degr,
20\degr, 37\degr, and -23\degr. For cells centered at 1.4 kpc and 2.0
kpc, there are fewer than 5 background stars which is below our threshhold
for analysis as discussed in \S\ 3.5.3.  Although there is thus significant
variation with distance, in all five cases the field direction is closer
to being perpendicular to the Galactic plane than to being parallel
to the plane.

We changed the size of the cell centered on NGC 6334 to examine the effect
on mean field direction. Shrinking any of the three dimensions to half
the original size will not change the mean field direction by more than
8\degr. Increasing the size along either the line-of-sight or
latitude directions will not change the mean field direction by more than 15\degr. 
Finally, if we increase the dimension along
the longitudinal direction by a factor of two, the o.p. drops from 0.41 to 0.25, leading to rejection
of the cell (see \S\ 3.5.3). 

\subsection{Comparisons with simulations} 

\subsubsection{Estimating magnetic field strength from the degree of field disorder }

Better constraints on magnetic field strengths in molecular clouds are
needed in order to constrain theories of star formation (Crutcher 2004).  In
principle, the method of Chandrasekhar \&\ Fermi (1953) can be used to derive
the field strength from the dispersion in measured submillimeter
polarization directions (Ostriker, Gammie, \&\ Stone 2001, Heitsch et al.
2001).  In practice the effects of beam dilution complicate this issue
(Heitsch et al. 2001, Houde 2004), and this problem is especially severe for
SPARO due to the 4\arcmin\ beamsize.  For this reason, we will instead make
a direct comparison of the degree of disorder in field angle seen our measurements with
that seen in three dimensional MHD turbulence simulations of GMCs presented
by Ostriker et al.\ (2001), after first smoothing the latter to the coarse
resolution of SPARO.  We will again use the order parameter (o.p.; \S\ 3.3) to
quantify the disorder in magnetic field direction.

Ostriker et al.\ (2001) use their numerical simulations to follow the time evolution of
initially smooth, self-gravitating, isothermal gas.  The initial velocity
field corresponds to Komolgorov-type turbulence with Mach number equal to 14. The
initial magnetic field is uniform, but three models are considered with
three different field strengths corresponding to $\beta$ = 1, 0.1, and 0.01, where
$\beta$ = $c_{s}^2/v_{a}^2$, $c_{s}$ is the sound velocity, and $v_{a}$ is the Alfven velocity. During
the evolution, the turbulence decays and the Mach number drops. Snapshots of
density, velocity, and magnetic field structure are shown for various Mach
numbers.

Ostriker et al.\ (2001) give simulations of optical
polarization measurements for background stars observed through their simulated clouds. 
It is assumed
that all grains along the line of sight have equal polarization efficiency.  
Two such optical polarization maps are
given, both corresponding to M = 7, with $\beta$ = 1 and 0.01, respectively
(Figs.\ 22 and 23 of Ostriker et al. 2001).  The corresponding $\beta$ = 0.1 map is not shown, but the
authors note that it is very similar to the $\beta$ = 1 case. We have obtained
the $\beta$ = 0.1 map from E. Ostriker (private communication), and we average
the o.p. values obtained from these two maps and refer to the average values as
the o.p. for the ``$\beta \geq\ 0.1$ case''.

For grains aligned by a uniform magnetic field, the degree of optical polarization is linearly
proportional to column density, while the submillimeter polarization
magnitude has no dependence on column density (for low optical depth). This illustrates that it is 
the submillimeter polarized flux rather than the submillimeter polarization magnitude that 
will generally be proportional to the optical 
polarization magnitude. Thus, when we
bin optical polarization vectors to SPARO's resolution, we should combine them
as if they were polarized flux, using the method
of Stokes' parameters. In principle, we should combine 
Stokes' parameters derived from each simulation resolution element that lies within
a 4\arcmin\ SPARO beam, 
but Ostriker et al.\ (2001) only give optical polarization
vectors for a 12 $\times$ 12 grid of points.  However, this grid is dense enough 
to provide many optical vectors per SPARO beam (see below) so it should be
sufficient for reasonable estimates. 

The length L of each side of the cubical simulation box of Ostriker et al.\
(2001) is not well defined, but by assuming a temperature of 10 K and a mean
density of 100 $cm^{-3}$ they obtain L = 8 pc for the M = 7 simulations. For NGC
6334, where our beam size is approximately 2 pc, we thus obtain 16 simulated SPARO vectors
from each simulated cloud map (each vector is derived from binning nine simulated optical vectors).  
The resulting o.p.\ values are 0.97 and 0.15 for $\beta$
= 0.01 and $\beta \geq 0.1$, respectively.  For G333.6-0.2, we obtain only four simulated SPARO
vectors (each comes from binning 36 simulated optical vectors) and we find o.p.\ values 
of 0.98 and 0.35 for the low and high $\beta$ cases, respectively. G331.5-0.1 is too distant for meaningful comparisons.  

Our results for the Carina nebula are not suitable for comparison with the
turbulence simulations of Ostriker et al.\ (2001).  The reason is that
these authors simulated the inertial range of the turbulence, with a
Kolgomorov type energy spectrum over the entire 8 pc simulation cube. Our
observation of Carina, on the other hand, shows field structure determined
by the $\sim$ 10 pc scale curvature of the H\footnotesize II\ \normalsize bubbles (Fig.\ 5), not a
turbulent cascade in the inertial range.  The field structure is the
result of energy injection, with bubbles depositing energy directly into
the field on $\sim$ 10 pc scales.

The simulated turbulence maps correspond to the case where the initial field direction
is perpendicular to the line of sight, so the o.p. values we have
derived from the simulations are in fact upper limits.  Fig.\ 24 of Ostriker et al.\ (2001) shows
that the angle between the field and the line of sight has a dramatic effect
on the degree of disorder in the inferred field directions.

From our maps of NGC 6334 and G333.6-0.2, we obtain o.p.\ values of 0.73 and 0.80,
respectively, indicating that $\beta$ = 0.01 is preferred over $\beta \geq\ 0.1$.  The
product of $\beta$ with the square of the Mach number gives the ``turbulent beta'',
$\beta_{t}$, which provides a direct comparison of magnetic and turbulent energy
densities.  We see that our comparison strongly favors $\beta_{t}$ = 0.5 over $\beta_{t} \geq\ 5$,
implying that the magnetic energy density is comparable to the turbulent
energy density.

There are two main limitations to our comparison of SPARO data with
simulations.  First, there is a relatively small overlap between the smaller
spatial scales of the simulations and the larger scales of the SPARO data. 
Secondly, the assumption that all grains have equal polarizing efficiency is
probably wrong (Cho \&\ Lazarian 2005).  Despite these questions, it seems
difficult to reconcile our observations of reasonably uniform fields
with models having
very weak fields.

\subsubsection{Comparing fields inside GMCs with larger scale Galactic fields}
We showed in \S\ 3.3 that three of our four target GMCs have mean field directions, determined from 
SPARO data, that are within 15\degr\ of the Galactic plane.  As we  noted, the probability for
obtaining this result by pure chance is below 2\%.  This suggests a correlation between GMC fields
and the larger-scale Galactic field.  NGC 6334 is the outlier in our sample of four GMCs, having its mean 
field rotated 70\degr\ clockwise from the plane.  
In \S\ 3.5 we used the Heiles (2000) stellar polarization database to examine Galactic fields on
scales roughly corresponding to the GMC accumulation length ($\sim$ 400 pc).  We obtained mean
field directions for a set of 10-11 regions, one of which is centered on the closest of our four
targets, NGC 6334, while none of the other SPARO targets is close enough to obtain reliable
determinations of field direction from the Heiles database.  Within this sample of 10-11 regions,
we again find a tendency for the ``accumulation-scale field'' to be parallel to the Galactic plane
(Figs.\ 9 and 10) and we find that NGC 6334 is again an outlier in the distribution, having its
``accumulation-scale field'' more aligned with the Galactic latitude direction than with the
Galactic longitude direction.
We can explain both the tendency for fields in GMCs to align with the Galactic plane and also the
agreement of the field direction in NGC 6334 with the direction of the Galactic field local to NGC 6334 by
supposing that fields in GMCs are generally parallel to the  local Galactic fields.  

Is this effect seen in simulations of GMC formation?  MHD simulations of the process whereby gas uniformly spread through a galactic disk becomes concentrated into GMCs via self-gravitating instabilities have been 
carried out, for example, by Kim \&\ Ostriker (2001), and Kim, Ostriker, \&\ Stone (2003).  The work by 
Kim \&\ Ostriker (2001) included only two dimensions (thin disk limit), and the particular
instabilities that they studied for the Galactic disk (as opposed to the Galactic center) took too
long to develop.  They were thus judged to be unlikely candidates for GMC formation unless they
could be enhanced by additional agents.  Nevertheless, for purposes of comparison with our
observations we note that their model M23 (Fig. 13 of their paper) shows continuity and alignment
between fields inside dense clouds and those in the lower density surrounding medium, while their
model M10 (Fig. 12) shows no such correlation.  The main difference is the field strength.  

In principle, comparisons of such models with our results could test theories for GMC formation,
but in practice such comparisons may lead to ambiguous results, for present simulations.  The
problem is that the simulations seem to produce only clouds having angular momentum perpendicular
to the Galactic plane.  For the 2d simulations this is true by definition, but even in the 3d
simulations of Kim et al.\ (2003) we find that angular momentum of collapsing gas clouds 
is orthogonal to the plane, presumably because it is inherited from Galactic shear.  In this case
the predominant mode for field distortion is to shear the fields out into spiraling patterns, with
the spiral lying in the Galactic plane (Fig.\ 12 of Kim \&\ Ostriker 2001).  Unfortunately, because 
SPARO sightlines extending from the Sun to disk GMCs are parallel to the plane, the in-plane spiral 
field distortion patterns described above will be indistinguishable from the case of continuous,
ordered fields extending into GMCs with no or little change in direction.  

Could the correlation we observed between GMC fields and local Galactic fields be merely a
consequence of field twisting induced by clouds rotating preferentially about axes orthogonal to
the Galactic plane?  While we cannot rule this out, it is worth noting that observations of GMCs in 
M 33 (Rosolowsky et al. 2003) have failed to find a strong tendency for cloud rotational axes to be 
perpendicular to the disk of this spiral galaxy.  Thus it seems that the more likely
explanation for the correlations we have discovered is that Galactic fields do indeed pass into
GMCs with little change in direction.  When more realistic MHD simulations are available that
better match the recent observational data on cloud rotation axes, it will be interesting to
compare them with our observations of magnetic field strucuture.

\acknowledgments
This work was supported by NSF Award OPP-0130389.
We thank Eve Ostriker for the courtesy of the simulated polarimetry map and Roger Hildebrand and Ellen Zweibel for helpful comments.  





\clearpage



\begin{figure}
\epsscale{1.0}
\plotone{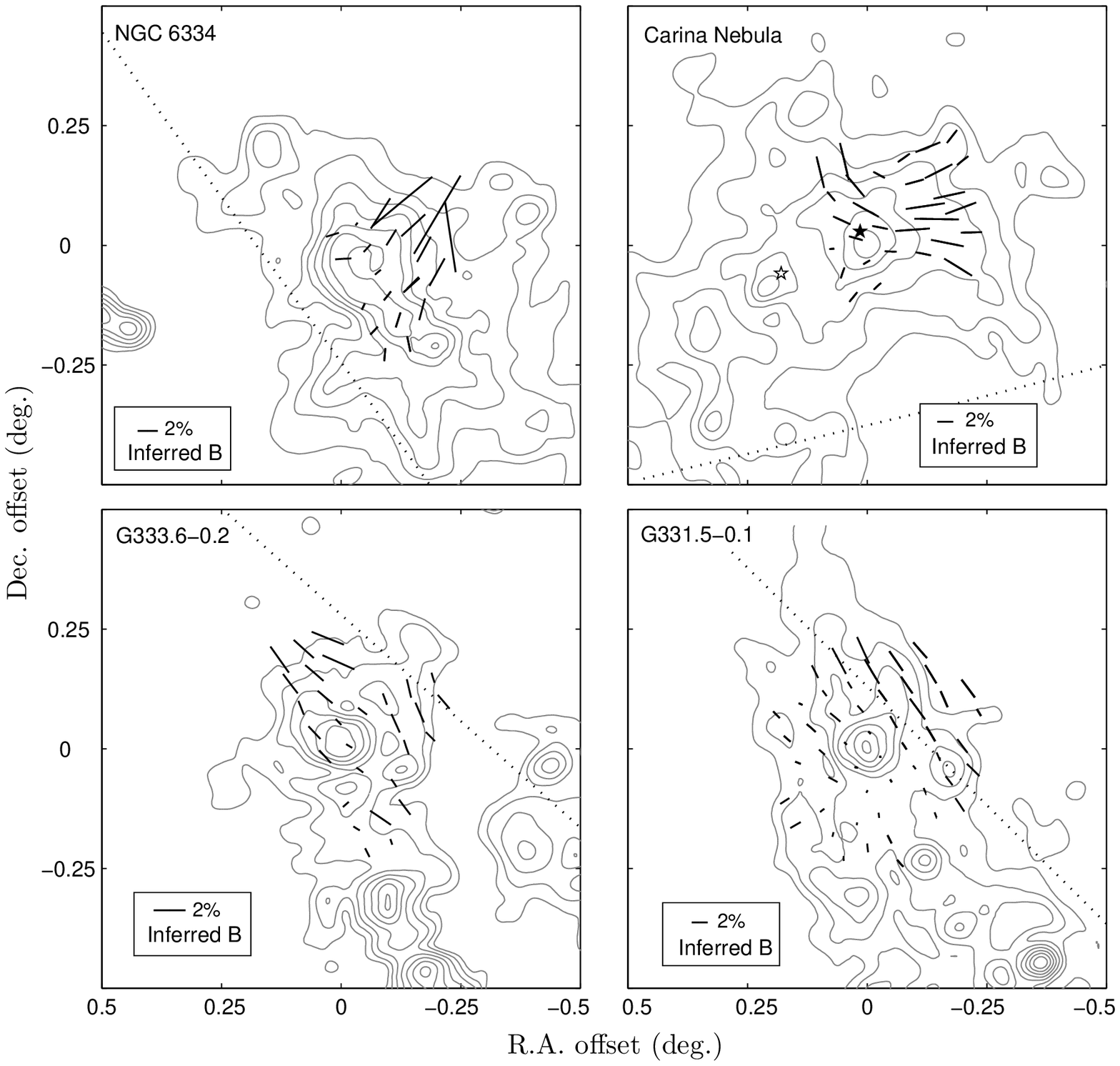} 
\caption{Results of SPARO 450 \micron\ polarimetric observations, shown together with the contours of 100 \micron\ HiRes IRAS data. The levels are 0.10, 0.18, 0.33, 0.60, 1.10, 2.00, and $3.65 \times 10^{4}$  MJy/sr.  
For each panel, the orientation of the Galactic plane is indicated as dotted lines at $b = 0.5\degr$\ (NGC 6334), $-1\degr$\ (Carina), and $0\degr$\ (G333.6 and G331.5), where $b$ is Galactic latitude. Each bar indicates a polarization detection with at least $2\sigma$ significance. Vectors are drawn parallel to the inferred magnetic field direction (i.e., perpendicular to the E-vector of the polarized emission). The length of each bar is proportional to the degree of polarization (see keys at lower left or right of each figure). 
The coordinate offsets are with respect to 
(17:20:53.7, -35:45:04) for NGC 6334, 
(10:43:20.0, -59:35:15) for Carina Nebula, 
(16:22:08.6, -50:06:59) for G333.6-0.2, and 
(16:12:10.3, -51:27:51) for G331.5-0.1 (J2000).}

\end{figure}

\clearpage


\begin{figure}
\epsscale{.9}
\plotone{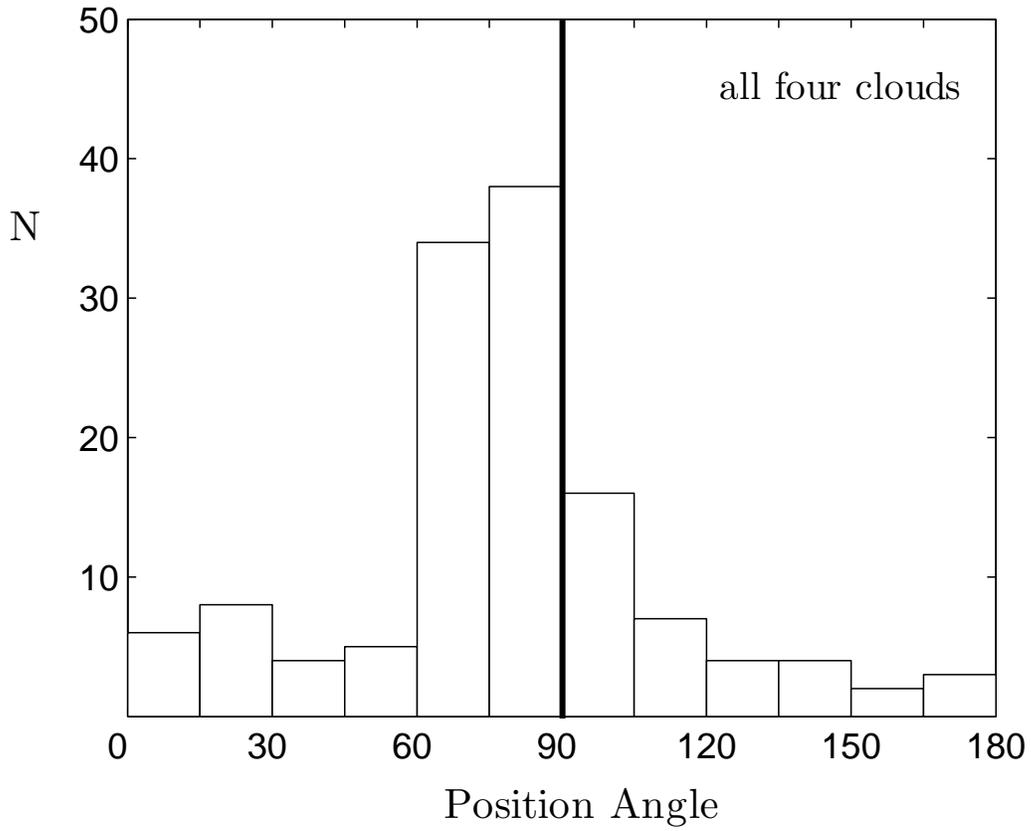} 
\caption{Histogram of magnetic field directions for all SPARO measurements (Tables 1--4), binned in 15\degr\ intervals. Position angles are measured from Galactic North-South, increasing counter-clockwise. }
\end{figure}


\begin{figure}
\includegraphics[angle=0,scale=.8]{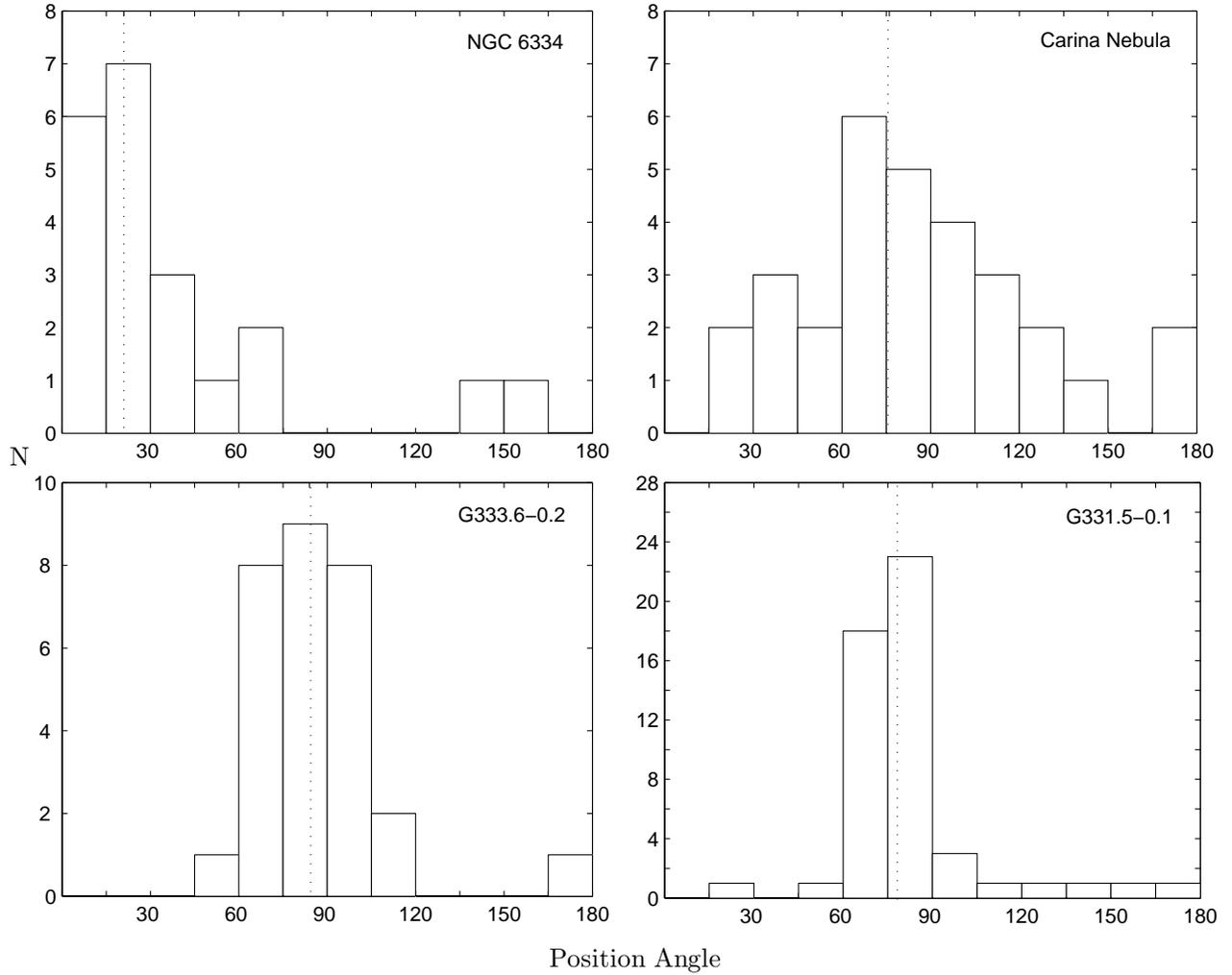} 
\caption{Histograms of magnetic field directions for all SPARO measurements, broken down by cloud. For each panel, a vertical dotted line 
shows the mean magnetic field direction for the cloud, computed as described in \S\ 3.3. Position angle is measured in Galactic coordinates,
as in Figure 2.}
\end{figure}

\begin{figure}
\epsscale{.9}
\plotone{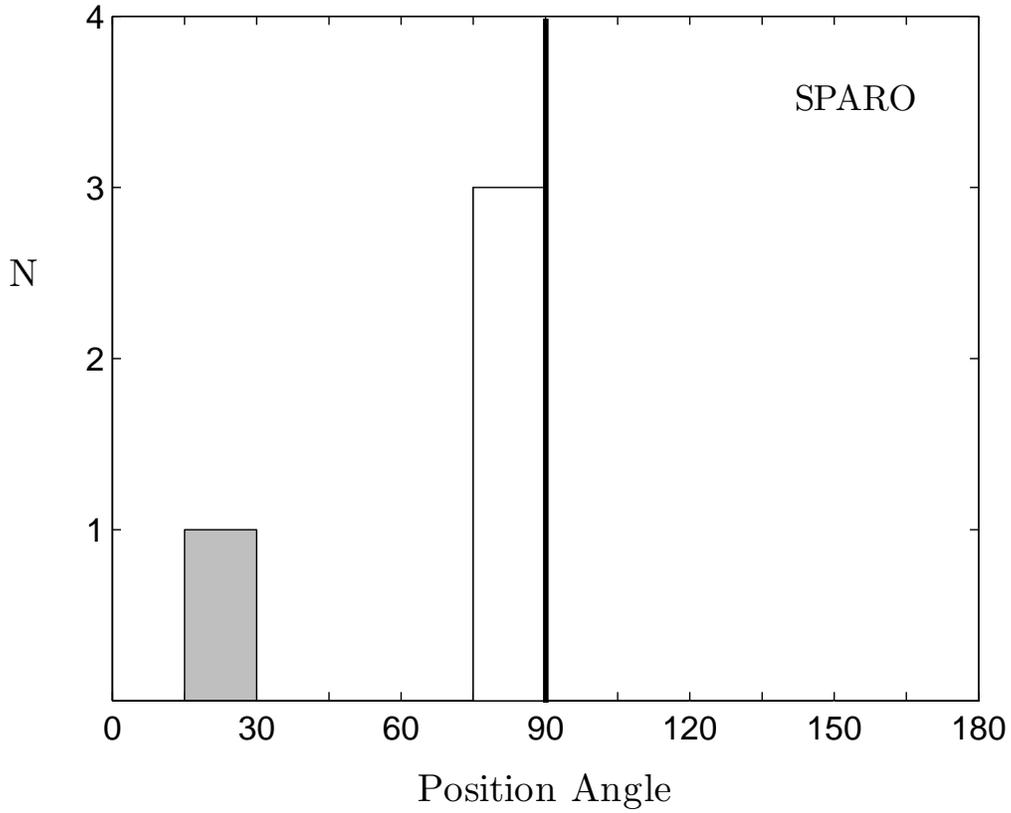} 
\caption{Histogram of the mean magnetic field direction for each GMC (\S\ 3.3). Position angle is measured in Galactic coordinates,
as in Figure 2.  Three out of the four clouds have mean fields that are within 15\degr\ of being
parallel to the Galactic disk.  NGC 6334 is the outlier, with a field rotated
away from the plane by $\sim 70\degr$.}
\end{figure}
\clearpage

\begin{figure}
\plotone{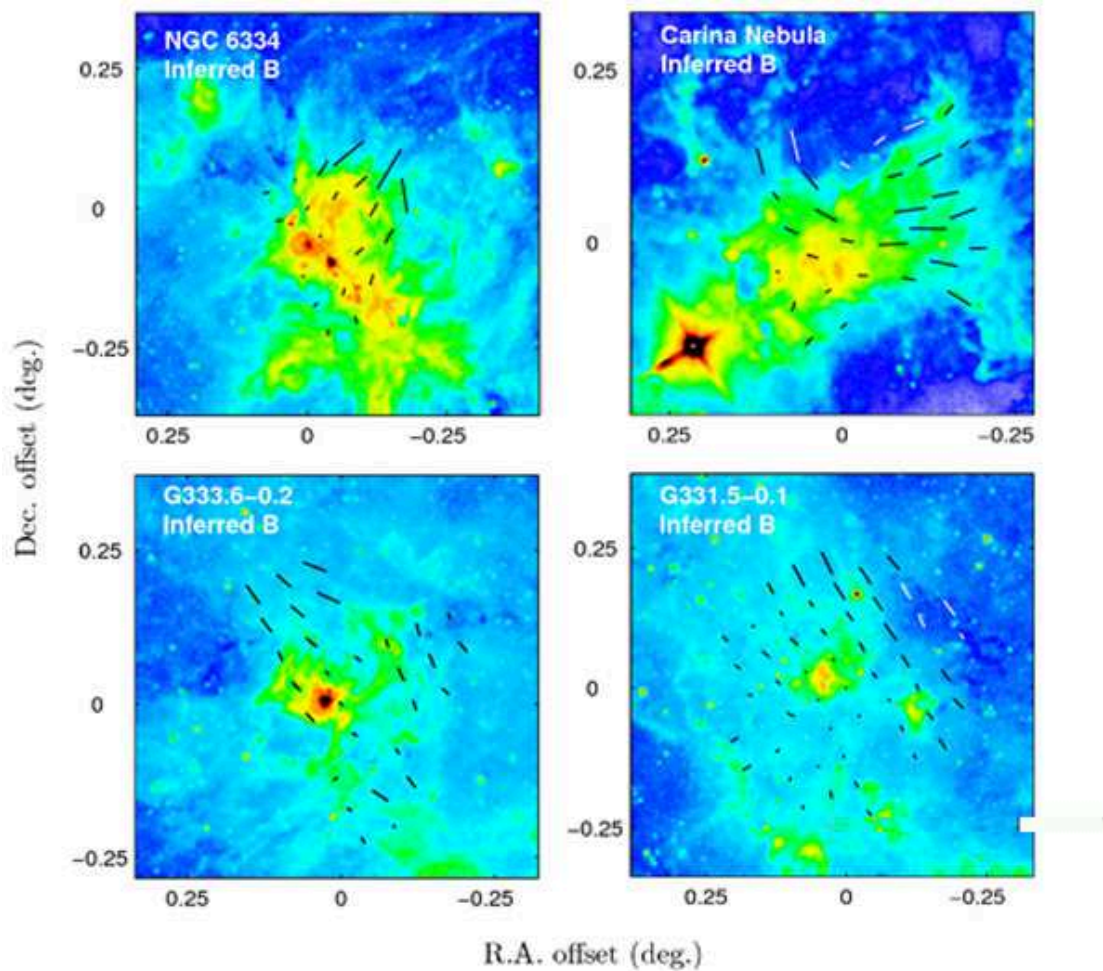}
\caption{Results of SPARO 450 \micron\ polarimetric observations, shown together with false color 8 \micron\ maps from the Midcourse Space Experiment (MSX). 
The vectors are drawn parallel to the inferred magnetic field direction and their lengths are proportional to the degree of polarization.
Note that in Carina the fields follow the curvature of the H\footnotesize II\ \normalsize bubbles. A similar effect 
is seen in G333.6-0.2 (\S\ 3.4).
The coordinate offsets are with respect to 
(17:20:51.0, -35:45:26) for NGC 6334, 
(10:43:15.6, -59:33:34) for the Carina Nebula, 
(16:22:03.4, -50:06:30) for G333.6-0.2, and 
(16:11:59.0, -51:28:40) for G331.5-0.1 (J2000).}
 
\small(Higher resolution figure can be downloaded at lennon.astro.northwestern.edu/f5.eps)

\end{figure}

\begin{figure}
\epsscale{0.9}
\plotone{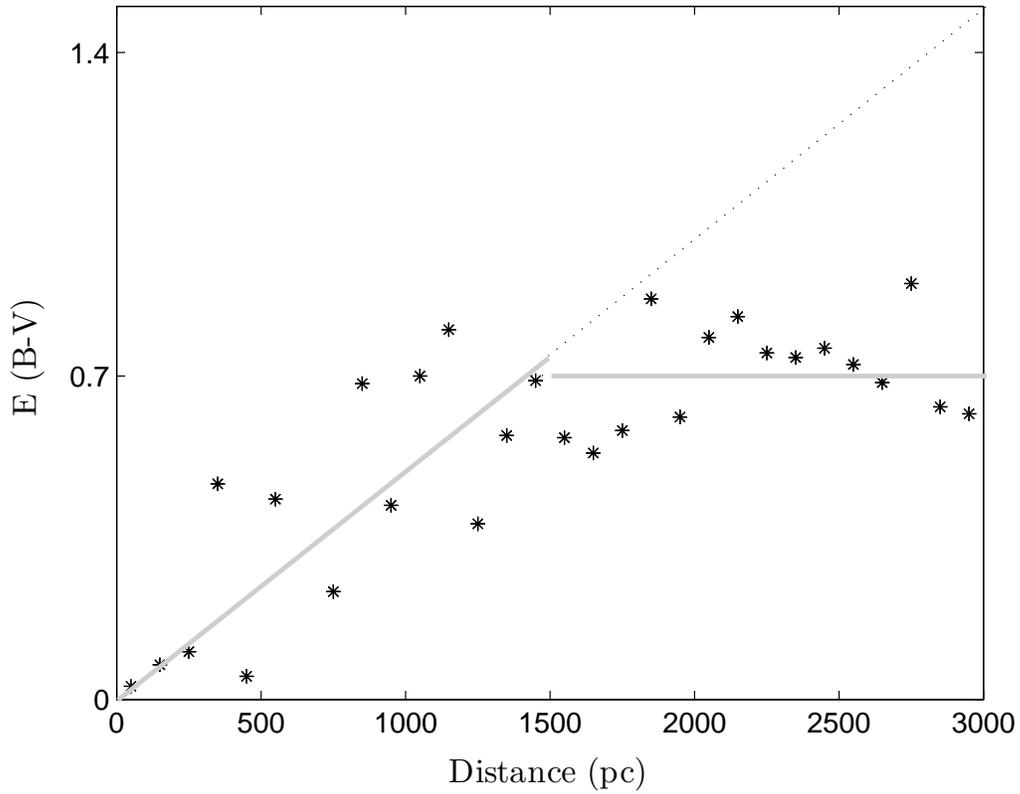} 
\caption{
Plot of mean selective extinction E(B-V) versus distance, for stars in the Helies (2000) database. We restricted the distance to the range shown
and Galactic latitude to $b = \pm 0.1\degr$. The data are binned in 100 pc intervals. 
The left solid line is a linear fit to the points having $d \leq 1500$ pc, forced to pass through the origin. The right solid line indicates the mean for points with $d > 1500$ pc, which is 0.7. The dotted line shows an extension of the left solid line. These lines are used to develop a correction for Malmquist bias (\S\ 3.5.4).}
\end{figure}

\begin{figure}
\epsscale{0.9}
\plotone{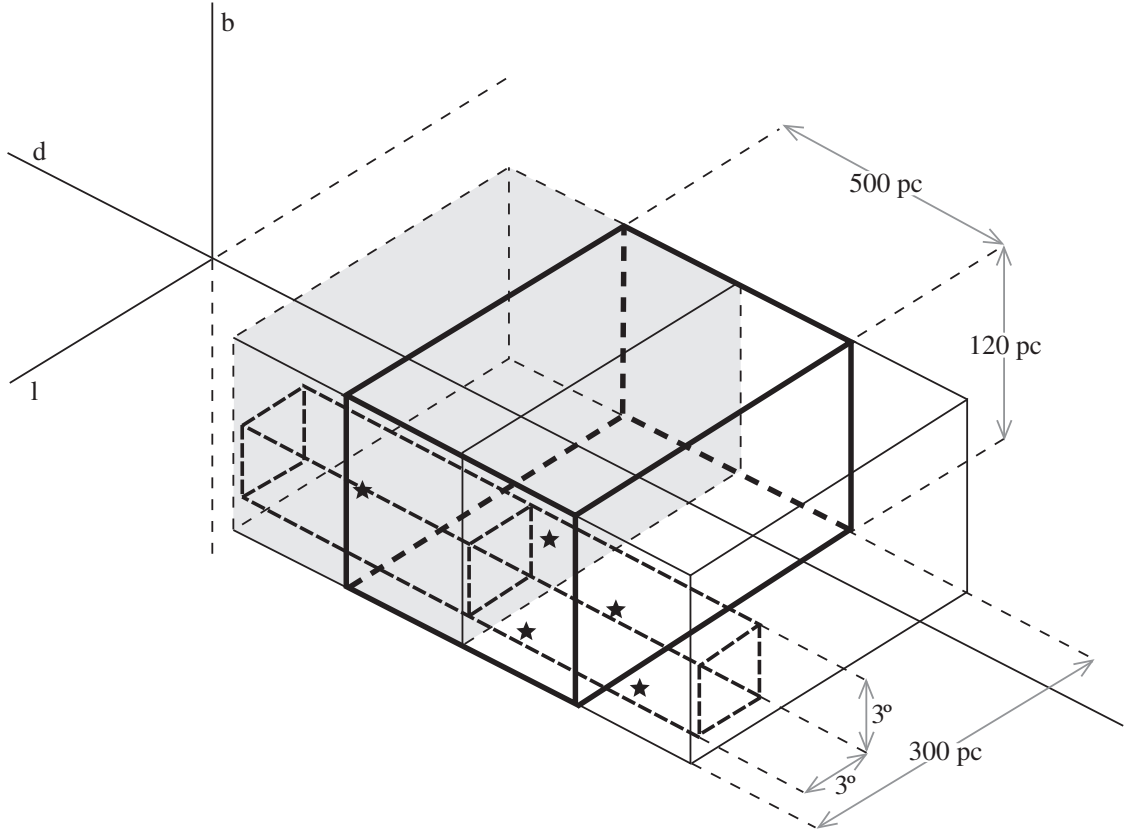} 
\caption{Main, foreground, and background cells, used for analysis of
stellar polarization data (\S\ 3.5.2).  Illustration is referenced to
($l$, $b$, $d$)-space.  Bold solid lines show the main cell (having
dimensions 500 pc $\times$ 300 pc $\times$ 120 pc) within which we want to estimate
the mean magnetic field angle.  Note that although the main cells are
cuboids in ($l$, $b$, $d$)-space, in real space they have slightly curved
faces and they tile the local region of the Galactic disk in three rings
(\S\ 3.5.2).  For each main cell, the corresponding background
(foreground)  cell is centered on the far (near) face of the main cell.  
The background cell is shaded.  Stars in the background (foreground) cell
are defined as background (foreground) stars.  For each background star
(e.g., Galactic coordinates [$l_0$, $b_0$]) several corresponding
foreground stars (restricted to [$l_0 \pm 1.5\degr$, $b_0 \pm 1.5\degr $])
are used for removing foreground polarization effects.  Star symbols show
one background star with four corresponding foreground stars.}
\end{figure}

\begin{figure}
\epsscale{1.2}
\plotone{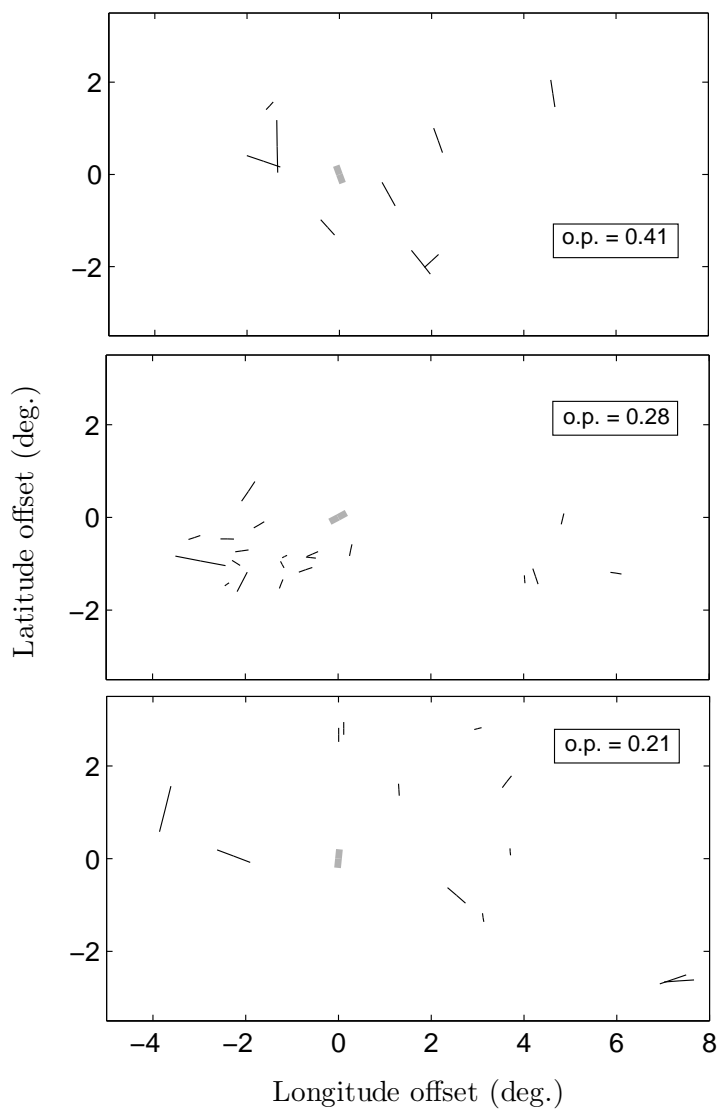} 
\caption{Three examples illustrating the use of the order parameter (o.p.). The thin, dark bars show optical polarization residues for three of the cells used in our analysis of optical polarization data (\S\ 3.5).  
Their orientations indicate the direction of the inferred magnetic field that is parallel to the angle of polarization, and their lengths are proportional to the degree of polarization.  
The key for these lengths is provided by the thicker bars in each panel, that are drawn with length corresponding to 1 \% polarization. Meanwhile, the orientation of each thick bar indicates the equal-weight-Stokes mean of the polarization
residues for each cell. Note that the greater the degree of uniformity in the directions of
the thin bars, the higher the corresponding o.p.\ value. The top panel corresponds to the cell centered on NGC 6334.}
\end{figure}

\begin{figure}
\epsscale{0.9}
\plotone{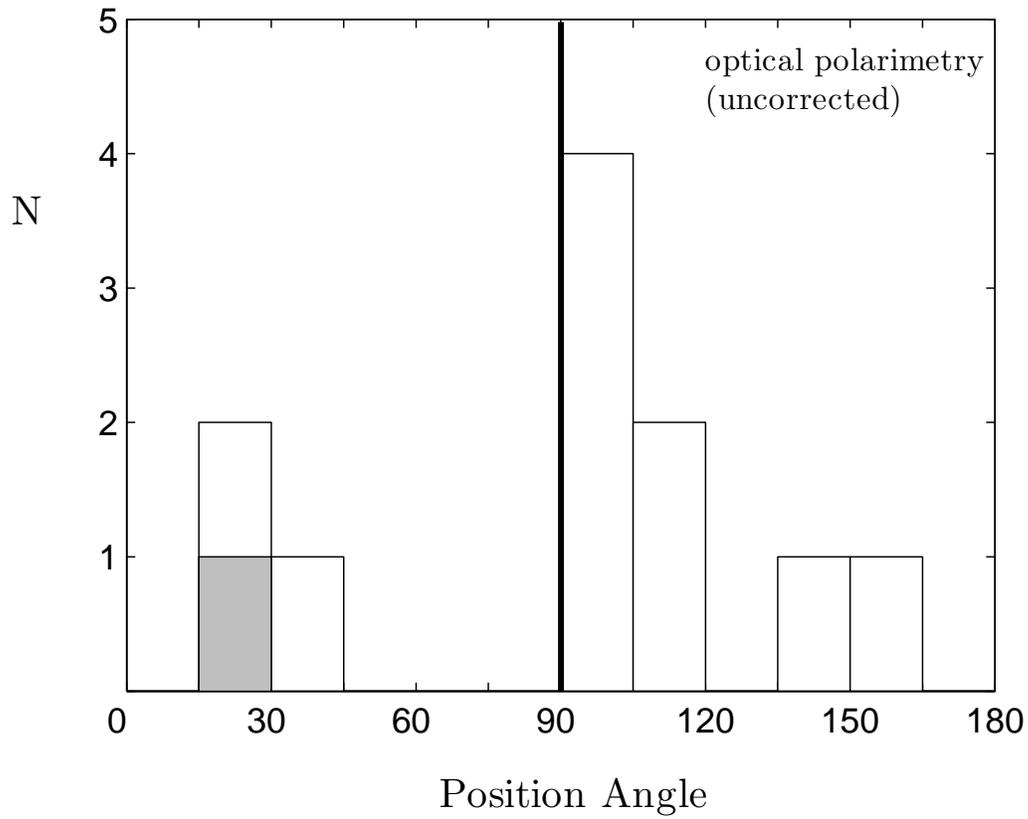} 
\caption{Histogram of magnetic field directions determined from optical polarimetry data, for 11 ``cells'' 
in the Galaxy (\S\ 3.5).  The shaded box corresponds to the cell centered on NGC 6334.  The data have
not been corrected for Malmquist bias. Position angle is measured in Galactic coordinates,
as in Figure 2.}
\end{figure}

\begin{figure}
\epsscale{0.9}
\plotone{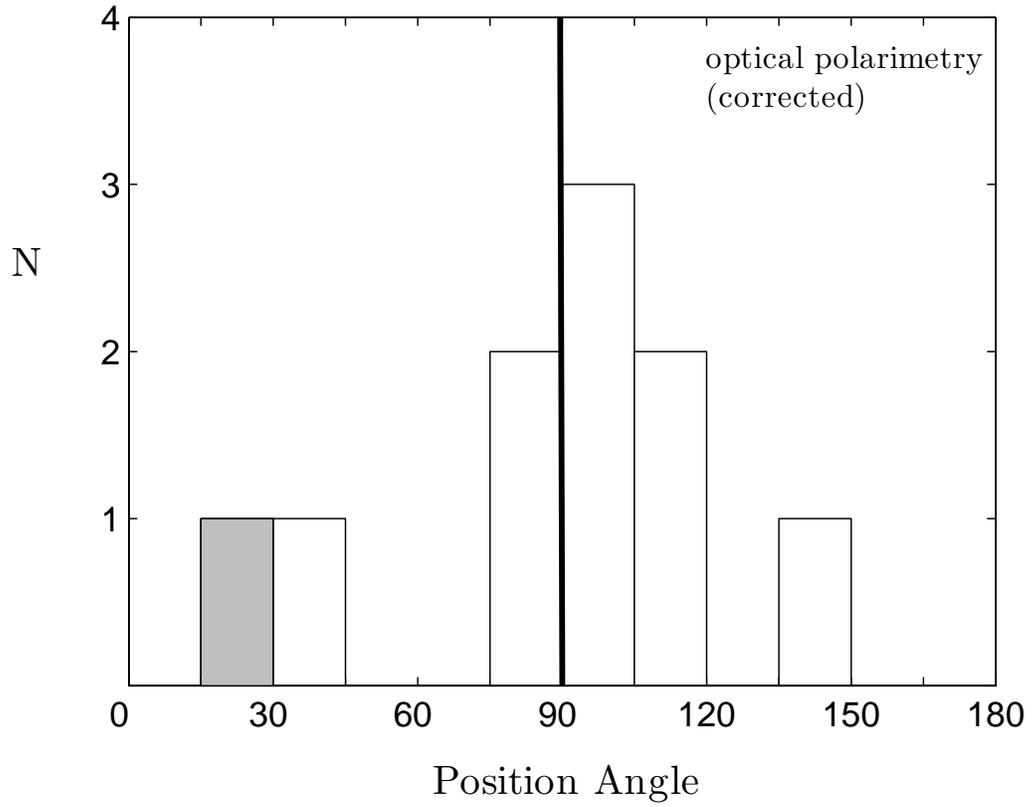} 
\caption{Similar to Figure 9, but with a correction for Malmquist bias applied. Again, the
cell corresponding to NGC 6334 is denoted by the shaded box.}
\end{figure}


\clearpage

\begin{deluxetable}{cccccc}
\tablewidth{0pt}
\tablecaption{NGC 6334 polarization results}
\tablehead{
\colhead{$\Delta\alpha$\tablenotemark{a}}           & \colhead{$\Delta\delta$\tablenotemark{a}}      &
\colhead{$P(\%)$}          & \colhead{$\sigma_{P}$}  &
\colhead{$\phi$\tablenotemark{b}}          & \colhead{$\sigma_{\phi}$}    }
\startdata
   3.01	&	-1.34&	1.8	&	0.16	&	3.0	&	2.6	\\
4.36	&	1.67&	1.49	&	0.24	&	16.6	&	4.7	\\
-1.34	&	-3.01&	0.84	&	0.12	&	40.1	&	4.2	\\
0	&	0    &	1.16	&	0.05	&	47.4	&	1.3	\\
1.34	&	3.01 &	0.38	&	0.13	&	55.5	&	9.8	\\
-3.01	&	1.34&	2.05	&	0.16	&	58.3	&	2.3	\\
-1.67	&	4.36&	3.99	&	0.49	&	57.3	&	3.5	\\
-7.36	&	-0.34&	3.18	&	1.18	&	61.3	&	10.6	\\
-6.02	&	2.68&	3.59	&	1.26	&	42.4	&	10.0	\\
-4.67	&	5.69&	8.56	&	3.24	&	39.4	&	10.9	\\
-9.03	&	4.02&	10.13	&	4.46	&	59.2	&	12.6	\\
-2.35	&	-13.39&	1.43	&	0.27	&	84.4	&	5.4	\\
-1.01	&	-10.37&	1.18	&	0.35	&	46.8	&	8.6	\\
0.34	&	-7.36&	0.75	&	0.25	&	65.1	&	9.5	\\
-5.36	&	-12.04&	1.76	&	0.17	&	101.8	&	2.7	\\
-4.02	&	-9.03&	1.72	&	0.21	&	72.4	&	3.5	\\
-2.68	&	-6.02&	1.24	&	0.21	&	51.9	&	4.9	\\
-7.03	&	-7.69&	2.49	&	0.23	&	75.0	&	2.6	\\
-5.69	&	-4.67&	2.37	&	0.58	&	42.0	&	7.0	\\
-8.7	&	-3.33&	3.49	&	0.9	&	61.1	&	7.4	\\
-10.37	&	1.03&	7.85	&	2.36	&	98.6	&	8.6	\\

\enddata
\tablenotetext{a}{Offsets in Right Ascension and Declination are
measured in arcminutes relative to the J2000 position (17h20m51.0s -35d45m26s).}

\tablenotetext{b}{$\phi$ is the angle of the E-vector of the polarized
radiation, measured in degrees from north-south, increasing
counterclockwise.}
\end{deluxetable}

\clearpage
\begin{deluxetable}{cccccc}
\tablewidth{0pt}
\tablecaption{Carina Nebula polarization results}
\tablehead{
\colhead{$\Delta\alpha$\tablenotemark{a}}           & \colhead{$\Delta\delta$\tablenotemark{a}}      &
\colhead{$P(\%)$}          & \colhead{$\sigma_{P}$}  &
\colhead{$\phi$}          & \colhead{$\sigma_{\phi}$}    }
\startdata
4.36	&	1.67	&	1.74	&	0.41	&	50.6	&	6.7	\\
5.71	&	4.68	&	1.61	&	0.27	&	71.2	&	4.8	\\
7.05	&	7.7	&	0.71	&	0.3	&	3.6	&	11.9	\\
1.34	&	3.01	&	1.4	&	0.36	&	42.4	&	7.3	\\
2.68	&	6.02	&	1.12	&	0.17	&	30.7	&	4.3	\\
4.02	&	9.03	&	1.98	&	0.23	&	163.9	&	3.4	\\
-0.33	&	7.38	&	1.66	&	0.21	&	177.2	&	3.6	\\
-6.02	&	2.68	&	5.45	&	0.41	&	8.5	&	2.2	\\
-4.68	&	5.71	&	2.67	&	0.34	&	15.4	&	3.6	\\
-3.34	&	8.72	&	2.04	&	0.41	&	35.2	&	5.8	\\
-9.03	&	4.02	&	4.46	&	0.44	&	12.2	&	2.8	\\
-7.7	&	7.05	&	4.28	&	0.45	&	26.7	&	3.0	\\
-6.35	&	10.06	&	3.67	&	0.46	&	21.7	&	3.6	\\
-10.7	&	8.38	&	1.99	&	0.76	&	37.7	&	10.9	\\
-9.36	&	11.37	&	2.27	&	0.87	&	52.2	&	11.1	\\
3.34	&	-8.72	&	1.74	&	0.41	&	50.6	&	6.7	\\
4.68	&	-5.71	&	1.61	&	0.27	&	71.2	&	4.8	\\
6.02	&	-2.68	&	0.71	&	0.3	&	3.6	&	11.9	\\
0.33	&	-7.38	&	1.4	&	0.36	&	42.4	&	7.3	\\
1.67	&	-4.36	&	1.12	&	0.17	&	30.7	&	4.3	\\
3.01	&	-1.34	&	1.98	&	0.23	&	163.9	&	3.4	\\
-1.34	&	-3.01	&	1.66	&	0.21	&	177.2	&	3.6	\\
0	&	0	&	2.18	&	0.12	&	169.1	&	1.6	\\
-5.71	&	-4.68	&	2.2	&	0.5	&	168.8	&	6.5	\\
-4.36	&	-1.67	&	4.79	&	0.31	&	4.3	&	1.8	\\
-10.06	&	-6.53	&	4.5	&	0.93	&	148.2	&	5.9	\\
-8.72	&	-3.34	&	4.6	&	0.57	&	168.7	&	3.5	\\
-7.38	&	-0.33	&	6.14	&	0.51	&	179.1	&	2.4	\\
-11.73	&	-2	&	2.96	&	0.7	&	1.4	&	6.8	\\
-10.39	&	1.02	&	4.56	&	0.62	&	20.4	&	3.9	\\

\enddata
\tablenotetext{a}{Offsets in Right Ascension and Declination are
measured in arcminutes relative to the J2000 position (10h43m15.6s -59d33m34s).}
\end{deluxetable}


\clearpage
\begin{deluxetable}{cccccc}
\tablewidth{0pt}
\tablecaption{G333.6-0.2 polarization results}
\tablehead{
\colhead{$\Delta\alpha$\tablenotemark{a}}           & \colhead{$\Delta\delta$\tablenotemark{a}}      &
\colhead{$P(\%)$}          & \colhead{$\sigma_{P}$}  &
\colhead{$\phi$}          & \colhead{$\sigma_{\phi}$}    }
\startdata
5.71	&	4.68	&	1.04	&	0.25	&	111.3	&	7.0	\\
7.05	&	7.7	&	1.69	&	0.21	&	127.2	&	3.5	\\
8.38	&	10.7	&	2.21	&	0.22	&	125.2	&	2.9	\\
2.68	&	6.02	&	1.35	&	0.29	&	139.9	&	6.1	\\
4.02	&	9.03	&	1.67	&	0.29	&	139.2	&	5.0	\\
5.36	&	12.04	&	1.87	&	0.27	&	138.1	&	4.2	\\
1.02	&	10.39	&	2.41	&	0.52	&	156.1	&	6.2	\\
2.35	&	13.39	&	2.4	&	0.48	&	158.3	&	5.7	\\
3.01	&	-1.34	&	1.2	&	0.19	&	132.8	&	4.6	\\
4.36	&	1.67	&	1.2	&	0.18	&	133.4	&	4.4	\\
-1.34	&	-3.01	&	0.57	&	0.12	&	144.2	&	5.8	\\
0	&	0	&	0.49	&	0.07	&	59.9	&	93.9	\\
1.34	&	3.01	&	0.57	&	0.11	&	133.0	&	5.3	\\
-1.67	&	4.36	&	0.78	&	0.2	&	142.5	&	7.5	\\
-7.38	&	-0.33	&	1.06	&	0.14	&	108.4	&	3.9	\\
-6.02	&	2.68	&	1.41	&	0.14	&	114.6	&	2.8	\\
-4.68	&	5.71	&	0.83	&	0.28	&	110.2	&	9.7	\\
-10.39	&	1.02	&	0.89	&	0.23	&	135.0	&	7.6	\\
-9.03	&	4.02	&	1.5	&	0.3	&	113.4	&	5.7	\\
-7.7	&	7.05	&	1.32	&	0.25	&	103.3	&	5.4	\\
-12.04	&	5.36	&	1.25	&	0.39	&	130.7	&	8.9	\\
-10.7	&	8.38	&	0.72	&	0.26	&	107.7	&	10.2	\\
-2.35	&	-13.39	&	0.68	&	0.13	&	116.9	&	5.6	\\
-1.02	&	-10.39	&	0.52	&	0.24	&	148.1	&	13.4	\\
0.33	&	-7.38	&	0.53	&	0.23	&	40.3	&	12.2	\\
-5.36	&	-12.04	&	0.4	&	0.17	&	107.4	&	12.0	\\
-4.02	&	-9.03	&	1.74	&	0.4	&	145.9	&	6.6	\\
-7.05	&	-7.7	&	1.32	&	0.35	&	127.0	&	7.6	\\
-5.71	&	-4.68	&	0.65	&	0.15	&	124.2	&	6.5	\\

\enddata
\tablenotetext{a}{Offsets in Right Ascension and Declination are
measured in arcminutes relative to the J2000 position (16h22m03.4s -50d06m30s).}
\end{deluxetable}

\clearpage
\begin{deluxetable}{cccccc}
\tablewidth{0pt}
\tablecaption{G331.5-0.1 polarization results}
\tablehead{
\colhead{$\Delta\alpha$\tablenotemark{a}}           & \colhead{$\Delta\delta$\tablenotemark{a}}      &
\colhead{$P(\%)$}          & \colhead{$\sigma_{P}$}  &
\colhead{$\phi$}          & \colhead{$\sigma_{\phi}$}    }
\startdata
1.67	&	-4.36	&	0.4	&	0.07	&	62.2	&	4.7	\\
3.01	&	-1.34	&	0.48	&	0.04	&	167.3	&	2.5	\\
4.36	&	1.67	&	0.47	&	0.04	&	133.1	&	2.6	\\
-1.34	&	-3.01	&	0.44	&	0.07	&	120.8	&	4.6	\\
0	&	0	&	0.3	&	0.05	&	5.8	&	4.7	\\
1.34	&	3.01	&	0.46	&	0.04	&	132.4	&	2.8	\\
-3.01	&	1.34	&	1.14	&	0.11	&	123.2	&	2.8	\\
-1.67	&	4.36	&	1.87	&	0.11	&	119.2	&	1.8	\\
5.71	&	4.68	&	1.37	&	0.14	&	137.5	&	2.8	\\
7.05	&	7.7	&	1.12	&	0.19	&	121.3	&	4.8	\\
8.38	&	10.7	&	1.97	&	0.24	&	116.3	&	3.5	\\
2.68	&	6.02	&	1.26	&	0.11	&	129.9	&	2.4	\\
4.02	&	9.03	&	1.06	&	0.25	&	121.8	&	6.6	\\
5.36	&	12.04	&	4.07	&	0.59	&	117.9	&	4.2	\\
1.02	&	10.39	&	3.32	&	0.38	&	122.3	&	3.3	\\
2.35	&	13.39	&	4.02	&	0.6	&	114.6	&	4.3	\\
10.7	&	-8.38	&	1.65	&	0.55	&	30.3	&	9.5	\\
12.04	&	-5.36	&	1.5	&	0.56	&	33.0	&	10.6	\\
6.02	&	-2.68	&	1.43	&	0.29	&	151.6	&	5.8	\\
-6.02	&	2.68	&	2.4	&	0.38	&	120.6	&	4.6	\\
-7.7	&	7.05	&	2.45	&	0.77	&	113.8	&	9.1	\\
-12.04	&	5.36	&	1.14	&	0.44	&	120.2	&	11.0	\\
-10.7	&	8.38	&	3.11	&	0.84	&	127.0	&	7.7	\\
-2.35	&	13.39	&	1.23	&	0.37	&	132.2	&	8.6	\\
-1.02	&	10.39	&	1.23	&	0.29	&	116.0	&	6.7	\\
0.33	&	-7.38	&	0.69	&	0.3	&	94.8	&	12.7	\\
10.39	&	-1.02	&	0.81	&	0.29	&	158.7	&	10.2	\\
11.73	&	2	&	1.23	&	0.31	&	138.8	&	7.3	\\
13.08	&	5.01	&	1.29	&	0.26	&	133.5	&	5.9	\\
7.38	&	0.33	&	3.55	&	0.42	&	125.7	&	3.4	\\
8.72	&	3.34	&	2.81	&	0.43	&	124.4	&	4.4	\\
10.06	&	6.35	&	3.27	&	0.45	&	124.9	&	3.9	\\
-0.33	&	7.38	&	1.94	&	0.23	&	125.6	&	3.4	\\
-4.68	&	5.71	&	3.55	&	0.42	&	125.7	&	3.4	\\
-3.34	&	8.72	&	2.81	&	0.43	&	124.4	&	4.4	\\
-2	&	11.73	&	3.27	&	0.45	&	124.9	&	3.9	\\
-6.35	&	10.06	&	2.94	&	0.44	&	121.4	&	4.3	\\
-5.01	&	13.08	&	2.88	&	0.76	&	131.0	&	7.6	\\
5.01	&	-13.08	&	0.91	&	0.42	&	110.3	&	13.2	\\
6.35	&	-10.06	&	1.04	&	0.26	&	114.1	&	7.2	\\
7.7	&	-7.05	&	0.62	&	0.25	&	127.6	&	11.6	\\
2	&	-11.73	&	1.54	&	0.34	&	128.1	&	6.4	\\
3.34	&	-8.72	&	1.26	&	0.25	&	112.4	&	5.8	\\
-7.05	&	-7.7	&	0.91	&	0.42	&	110.3	&	13.2	\\
-5.71	&	-4.68	&	1.04	&	0.26	&	114.1	&	7.2	\\
-4.36	&	-1.67	&	0.62	&	0.25	&	127.6	&	11.6	\\
-10.06	&	-6.35	&	2.22	&	0.52	&	123.2	&	6.7	\\
-8.72	&	-3.34	&	1.54	&	0.34	&	128.1	&	6.4	\\
-7.38	&	-0.33	&	1.26	&	0.25	&	112.4	&	5.8	\\
-11.73	&	-2	&	2.42	&	0.48	&	135.4	&	5.6	\\
-10.39	&	1.02	&	2.35	&	0.46	&	126.5	&	5.6	\\

\enddata
\tablenotetext{a}{Offsets in Right Ascension and Declination are
measured in arcminutes relative to the J2000 position (16h11m59.0s -51d28m40s).}
\end{deluxetable}

\end{document}